\documentclass[a4paper,11pt]{article}
\pdfoutput=1 

\usepackage{jheppub} 
                     
\usepackage{mytikz}

\usepackage{float}		
\usepackage{graphicx}		
\usepackage{subcaption}

\newcommand{\tr}{\text{tr}}

\DeclareMathOperator\arctanh{arctanh}

\newcommand{\be}{\begin{equation}}
\newcommand{\ee}{\end{equation}}
\newcommand{\bea}{\begin{eqnarray}}
\newcommand{\eea}{\end{eqnarray}}
\newcommand{\beas}{\begin{eqnarray*}}
\newcommand{\eeas}{\end{eqnarray*}}
\newcommand{\ba}{\begin{array}}
\newcommand{\ea}{\end{array}}

\title{Interpolating between multi-boundary wormholes and single-boundary geometries in holography}

\author[a]{Alex May,}
\author[a]{Mark Van Raamsdonk}

\affiliation[a]{Department of Physics and Astronomy, University of British Columbia
6224 Agricultural Road, Vancouver, B.C., V6T 1W9, Canada}

\emailAdd{may@phas.ubc.ca}
\emailAdd{mav@phas.ubc.ca}

\abstract{The recent paper 1809.01197 described how states of a holographic CFT can be approximated by states of a large collection of non-interacting BCFTs, such that the dual of the new system accurately approximates an arbitrarily large causal patch of the original geometry. In this paper, we first describe in more detail the geometries dual to such discrete BCFT systems, emphasizing that they are multi-boundary wormholes in which it is not possible to move causally between different asymptotic regions. By reintroducing couplings between the BCFTs in various ways, we show that the wormholes can be made traversable, giving an intermediate class of geometries that interpolate between the multi-boundary wormhole and the original geometry that it approximates.}

\begin{document} 
\maketitle
\flushbottom

\section{Introduction}

\begin{figure}
\begin{center}
\begin{subfigure}[b]{.45\textwidth}
\centering
\includegraphics[scale=0.2]{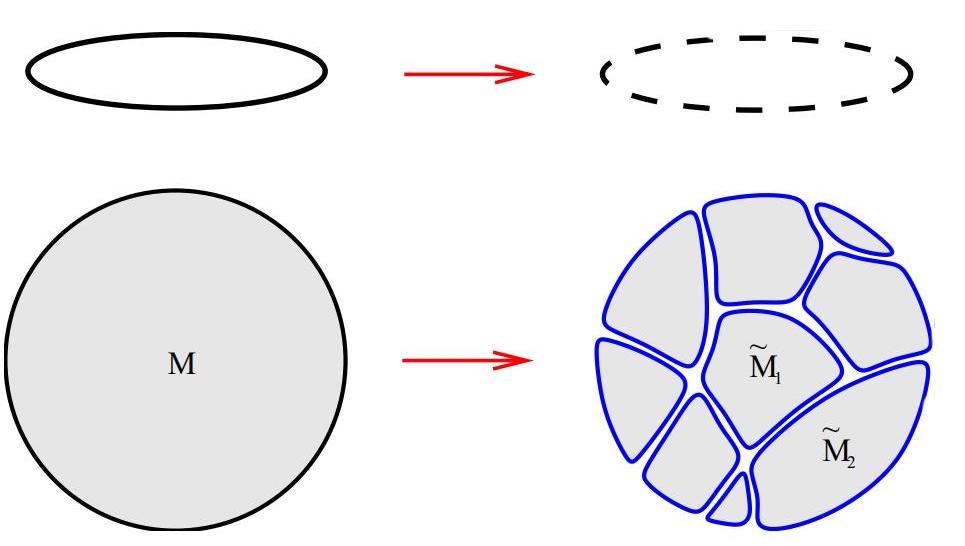}
\caption{}
\label{fig:reviewa}
\end{subfigure}
\hfill
\begin{subfigure}[b]{.45\textwidth}
\includegraphics[scale=0.19]{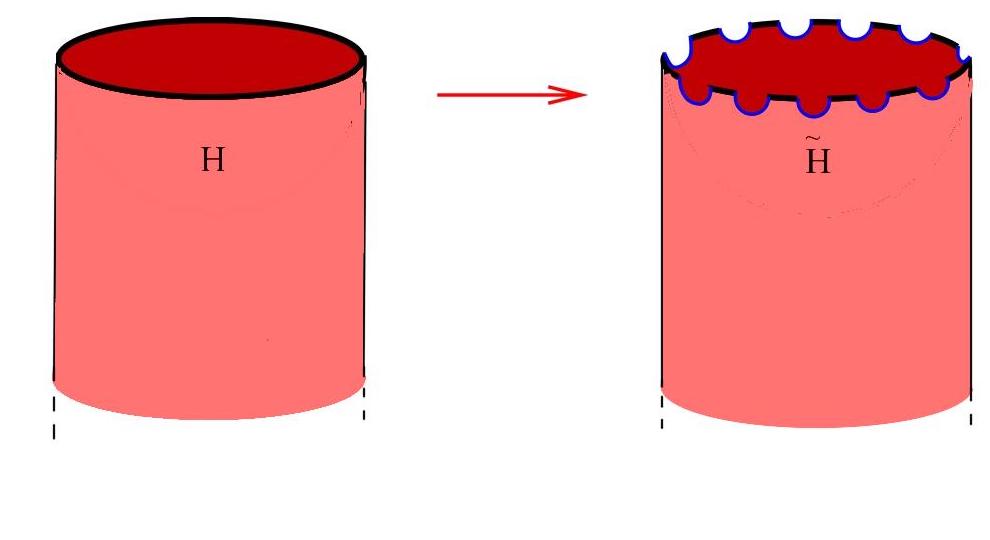}
\caption{}\label{fig:reviewb}
\end{subfigure}
\caption{(a) BCFT-bit systems approximating holographic 1+1 and 2+1 dimesional CFTs. (b) Path integral defining the state of BCFTs approximating the vacuum state of a holographic 1+1 dimensional CFT. In the limit where the modifications to the original path integral are small, the dual geometry faithfully approximates an arbitrarily large causal patch of the original AdS spacetime.}
\label{fig:review}
\end{center}
\end{figure}

Recent work in holographic approaches to quantum gravity suggests a direct relation between spacetime geometry and the quantum information (e.g. the structure of entanglement) stored in the fundamental degrees of freedom. An interesting aspect of this correspondence, emphasized recently in \cite{VanRaamsdonk:2018zws,VanRaamsdonk:2020ydg,simidzija2020holo} is that the precise nature of the holographic degrees of freedom may be relatively unimportant. Starting with a holographic CFT state encoding a particular spacetime, we can define a state of a completely different holographic CFT \cite{simidzija2020holo}, or of a collection of a large number of non-interacting BCFTs \cite{VanRaamsdonk:2018zws} (dubbed ``BCFT bits'' or ``BC-bits'') such that the new state is dual to a geometry that accurately approximates an arbitrarily large causal patch of the original spacetime (figure \ref{fig:review}). This is consistent with the idea that the encoded spacetime geometry and gravitational physics of the interior geometry is determined by the quantum information-theoretic properties of the state rather than which specific degrees of freedom the state is stored in.

In this paper, we investigate in more detail the spacetimes dual to holographic states approximated by discrete systems and describe various ways that we can interpolate between these geometries, which can be interpreted as non-traversable multi-boundary wormholes, and the original causally connected geometry. From the CFT point of view, the interpolation involves adding back interactions between the boundary degrees of freedom in the BCFTs in order to render the wormhole traversable. 


We begin in section \ref{sec:reviewofBCbits} by reviewing the BCFT-bit construction of \cite{VanRaamsdonk:2018zws}. We provide an alternative description via a ``discretization'' operator that maps from the Hilbert space of a holographic CFT to the Hilbert space of a collection of BCFTs.

In section \ref{sec:holographicduals}, we describe in more detail the geometries dual to the approximating BCFT-bit states. We emphasize that because there is no interaction between the components, the bulk geometries must have the property that it is not possible to travel between the asymptotic boundaries associated with different components. Thus, the geometries are non-traversable multi-boundary wormholes.\footnote{Here, each boundary component is asymptotically AdS but the boundary geometries have the topology of an interval/ball times time rather than a circle/sphere times time as in the more traditional wormholes of e.g. \cite{Balasubramanian:2014hda}.} Nevertheless, their interiors contain a causally connected region that approximates an arbitrarily large causal region of the original spacetime.

Making use of a simple bottom-up model for the gravity dual of BCFTs \cite{Karch2000,Takayanagi:2011zk}, we describe explicitly the approximated version of a pure $AdS^3$ geometry dual to the vacuum state of a holographic two-dimensional CFT. We show that the non-traversability is a result of the expansion of end-of-the-world branes that connect the boundary components, as shown in figure \ref{fig:expandingb}.

From the wormhole point of view, passing from the BCFT picture back to the original CFT amounts to rendering the wormholes traversable by coupling the degrees of freedom associated with the separate asymptotic regions in a very particular way. The main goal for the later sections of the paper is to understand various intermediate situations where the discrete BCFT systems are coupled together, but less strongly that in the original CFT. For various possible types of couplings, we will understand whether traversable geometries are obtained, and explore the details of the dual geometries.

\begin{figure}
\begin{center}
\begin{subfigure}[b]{0.45\textwidth}
\begin{center}
\begin{tikzpicture}
    
    \draw[thick,variable=\x,domain=0:180] (-3,0) -- (-1,0) plot ({-cos(\x)},{-sin(\x)}) -- (3,0);
    
    \draw[thick,variable=\x,domain=0:180,opacity=0.3,fill=red] (-3,0) -- (-1,0) plot ({-cos(\x)},{-sin(\x)}) -- (3,0) -- (3,-3) -- (-3,-3) -- (-3,0);
    
    \end{tikzpicture}
\end{center}
\caption{}
\label{fig:perturbedPIa}
\end{subfigure}
\hfill
\begin{subfigure}[b]{.45\textwidth}
\begin{center}
\begin{tikzpicture}
    
    \draw[thick,red] (-3,0) -- (-1,0);
    \draw[purple,thick,domain=0:180] plot ({-cos(\x)},{-sin(\x)});
    \draw[thick] (1,0) -- (3,0);
    
    \draw[thick,variable=\x,domain=0:180,opacity=0.3,fill=red] (-3,0) -- (-1,0) plot ({-cos(\x)},{-sin(\x)}) -- (3,0) -- (3,-3) -- (-3,-3) -- (-3,0);
    
    \draw[fill=blue,opacity=0.4,domain=0:180] (-1,0) -- (1,0) plot ({-cos(\x)},{-sin(\x)});
     
    \draw[thick,blue] (-1,0) -- (1,0);
     
    \node at (0,-2) {CFT$_1$};
    \node at (0,-0.5) {CFT$_2$};
     
\end{tikzpicture}

\end{center}
\caption{}
\label{fig:perturbedPIb}
\end{subfigure}
\begin{subfigure}[b]{.45\textwidth}
\begin{center}
\begin{tikzpicture}
    
    \draw[thick,red] (-3,0) -- (-1,0);
    \draw[purple,thick,domain=0:180] plot ({-cos(\x)},{-1.2*sin(\x)});
    \draw[purple,dashed,domain=0:180] plot ({-cos(\x)},{-1*sin(\x)});
    \draw[thick] (1,0) -- (3,0);
    
    \draw[thick,variable=\x,domain=0:180,opacity=0.3,fill=red] (-3,0) -- (-1,0) plot ({-cos(\x)},{-1.2*sin(\x)}) -- (3,0) -- (3,-3) -- (-3,-3) -- (-3,0);
    
    \draw[fill=blue,opacity=0.4,domain=0:180] (-1,0) -- (1,0) plot ({-cos(\x)},{-1.2*sin(\x)});
     
    \draw[thick,blue] (-1,0) -- (1,0);
     
    \node at (0,-2) {CFT$_1$};
    \node at (0,-0.5) {CFT$_2$};
     
\end{tikzpicture}
\end{center}
\caption{}
\label{fig:perturbedPIc}
\end{subfigure}
\hfill
\begin{subfigure}[b]{.45\textwidth}
\begin{center}
\begin{tikzpicture}

    \node at (0,1) {};
    
    \draw[gray] (-3,-3) -- (-3,0) -- (3,0) -- (3,-3) -- (-3,-3);
    
    \draw[gray,fill=red,opacity=0.3] (-3,-3) -- (-3,0) -- (-1,0) -- (-1,-3) -- (-3,-3);
    \draw[gray,fill=red,opacity=0.3] (3,-3) -- (3,0) -- (1,0) -- (1,-3) -- (3,-3);
    
    \draw[fill=blue,opacity=0.4] (-1,0) -- (1,0) -- (1,-3) -- (-1,-3) -- (-1,0);
    
    \draw[purple,thick] (-1,0) -- (-1,-3);
    \draw[purple,thick] (1,0) -- (1,-3);
    
    \node at (0,-1.5) {CFT$_2$};
    \node at (2,-1.5) {CFT$_1$};
    \node at (-2,-1.5) {CFT$_1$};
    
    \end{tikzpicture}
\end{center}
\caption{}
\label{fig:perturbedPId}
\end{subfigure}
\caption{(a) Path integral for the state of two BCFTs approximating the vacuum state of CFT${}_1$ (b) Path integral for an interface theory with an auxiliary CFT coupling the left and right BCFTs. The Lorentzian continuation of the associated Euclidean geometry remains non-traversable. (c) Deformation of the path-integral geometry leading to a traversable dual geometry. (d) Path-integral for the vacuum state of the interface CFT. The dual geometry is traversable, but in some cases, this requires crossing an interface brane.}
\label{fig:perturbedPI}
\end{center}
\end{figure}
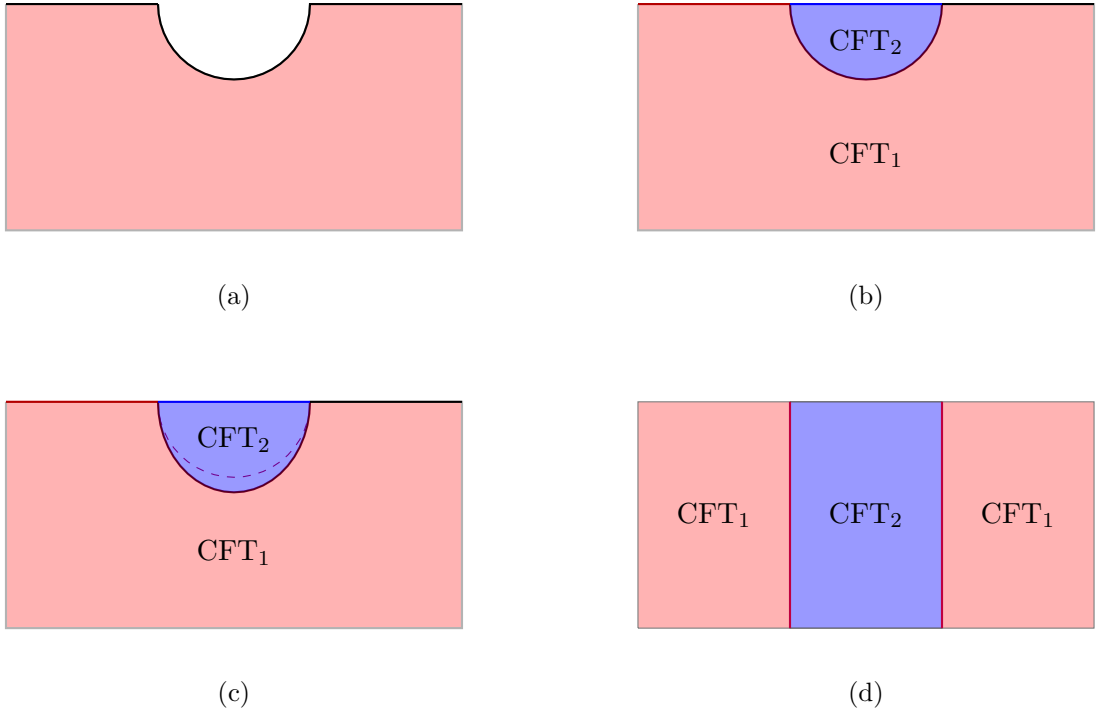

In these investigations, we simplify to the situation where the state of a holographic 1+1 dimensional CFT on $\mathbb{R}$ is approximated by a state of two BCFTs, each on a half space, i.e. by a theory obtained from the original one by removing a single interval, as shown in figure \ref{fig:perturbedPIa}. In this case, the simplest choice for the discretization operator acting on the original CFT vacuum state produces the thermofield double state of the two BCFTs. The corresponding dual geometry is reviewed in section \ref{sec:holographicduals}.

Starting from the initially decoupled system in the entangled state dual to a non-traversable geometry, we consider in section \ref{sec:doubletrace} coupling the two BCFTs by adding a relevant interaction between their boundaries:
\be
\label{GJW}
H \to H + f(t) {\psi}_R {\psi}_L \; .
\ee
Here ${\psi}_R$ and ${\psi}_L$ are CFT boundary operators with sufficiently low dimension. This is a boundary version of the coupling considered by Gao, Jafferis and Wall \cite{gao2017traversable} to render the wormhole in a two-sided Schwarzschild black hole traversable. We find similarly that such a coupling in our case leads to corrections that allow traversability. Even though the coupling involves only the boundary points, the geometry dual to the perturbed state can be traversed causally from any spatial location in one asymptotic region to any spatial location in the other asymptotic region. 

At least naively, the coupling (\ref{GJW}) with a single light operator leads to modifications to the geometry that appear at the quantum level. This can be understood by looking at the $N$-scaling of the modifications to correlators induced by the perturbation.\footnote{This analysis relies on an all-orders perturbative result; it is possible that non-perturbative effects could modify these expectations.} In section \ref{sec:intermediateCFT}, we consider a stronger type of coupling that affects the geometry at the classical level. Specifically, we couple neighboring boundaries via auxiliary degrees of freedom which take the form of an auxiliary CFT on an interval, which we allow in general to have a time-dependent size. The result is an interface theory where the ``gap'' in the BCFT picture is filled in with the degrees of freedom of a second CFT whose central charge we can control. This CFT is coupled to the left and right BCFTs at a pair of interfaces. In the limit where the central charges match, the interval size is fixed, and the interfaces are trivial, we recover the original CFT. However, we can also consider the central charge of the new CFT to be significantly smaller. 

Assuming that the new CFT also has a geometrical gravity dual, the dual bulk geometry now has two regions separated with different cosmological constants by an interface brane that emerges from the interface between CFTs at the boundary. In some cases (when the auxiliary degrees of freedom live on an interval that expands sufficiently rapidly), we find that the dual geometry remains non-traversable between the asymptotic regions associated with the left and right BCFTs. In other cases, the dual geometries include causal curves that connect the asymptotic regions associated with the left and right BCFTs, but these curves necessarily cross the interface brane. Thus, while the geometry is mathematically traversable, travelling from right to left is not possible for an observer made from the light fields in the effective gravity theory associated with our original CFT.  Finally, we may have situations where the geometry can be traversed without crossing the interface brane. We refer to such spacetimes as ``pleasantly traversable.''

As examples, we consider geometries obtained by the analytic continuation of the Euclidean geometries associated with the path integrals in figures \ref{fig:perturbedPIb} and \ref{fig:perturbedPIc}. For the path integral in figure \ref{fig:perturbedPIb} (where the doubled bra-ket path integral has rotational symmetry), the size of the auxiliary CFT interval in the Lorentzian picture increases with uniform acceleration and we find that the geometries are not traversable. However, even small perturbations such that the interface is a ``tall'' ellipse as in figure \ref{fig:perturbedPIc} give pleasantly traversable dual geometries. The Lorentzian geometries obtained by analytic continuation in this way have time-dependent interface trajectories in the CFT that are the analytic continuations of the Euclidean trajectories. The same initial states can be evolved with a interface CFT Hamiltonian where the interval size is static; in these cases, we argue that the dual geometries settle to static, traversable geometries, associated with the path integral shown in figure \ref{fig:perturbedPId}. These geometries may be pleasantly traversable or not, depending on the central charges and properties of the interface.

\section{Review of the BC-bit construction}\label{sec:reviewofBCbits}

We begin with a brief review of the construction in \cite{VanRaamsdonk:2018zws}, where states of a holographic CFT are approximated by states of a collection of BCFTs. 

Given a CFT on a spatial geometry $M$, we first choose a disconnected geometry $\{\tilde{M}_i\}$ whose pieces cover a large subset of the original geometry, as shown in figure \ref{fig:reviewa}. The BCFT-bit system is then defined on $\{\tilde{M}_i\}$ by choosing some boundary condition for the CFT on the boundaries of the pieces $\tilde{M}_i$.\footnote{To avoid confusion with the bulk-boundary terminology of AdS/CFT, we will refer to the boundaries of these pieces as CFT boundaries or \emph{edges}, $M$ or $\{\tilde{M}_i\}$ as the boundary, and the AdS spacetime as the bulk.} We can take this boundary condition to locally preserve conformal invariance, so that we have a BCFT associated with each individual piece.

Now, given a state $|\Psi \rangle$ of the original holographic CFT dual to some spacetime, we would like to approximate this state by a state of our collection of BCFTs. If $|\Psi \rangle$ is defined via a Euclidean path integral with sources (e.g. the semi-infinite cylinder path integral that defines the vacuum state), the approximated state is defined by introducing small boundary components to this path integral as shown in figure \ref{fig:reviewb}, so that it defines an entangled state of the BCFTs.

As explained in \cite{VanRaamsdonk:2018zws}, if these modifications to the path integral are small (for example, the gaps in figure \ref{fig:review} should be small compared to the remaining BCFT segments), the new dual geometry should accurately approximate an arbitrarily large causal patch of the original spacetime.\footnote{To see this in the two dimensional case on which we focus in this paper, note that the modifications to the doubled path integral from which the full geometry is deduced are circular boundaries whose radius goes to zero. In the limit where the circle becomes small, this region may be conformally mapped to a semi-infinite cylinder, which always prepares the vacuum state, implying the original operator approaches the identity. This is also apparent in the holographic model that we review below. There the circular boundaries introduced into the CFT path integral become the boundaries of end-of-the-world (ETW) branes. In the limit where the circles become small, these ETW branes have the geometry of a disk that is localized near the AdS boundary. In higher dimensions a similar bulk argument applies assuming the ETW brane model, but we do not have a CFT argument. Thus in higher dimensions we expect this to hold at least for appropriate choices of boundary condition.} More specifically, this causal patch is the domain of dependence of an arbitrarily large compact subset of the $t=0$ slice. 

\subsubsection*{The discretization operator}

Before proceeding, we give an alternative description of the approximation procedure that can be applied to general states of the CFT which are not necessarily described in terms of a Euclidean path integral.

\begin{figure}
    \centering
    \includegraphics[width = 40mm]{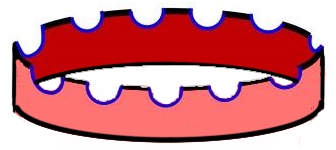}
    \caption{Euclidean path integral defining a ``discretization'' operator that maps from the Hilbert space of a CFT on $S^1$ to the Hilbert space of an associated BCFT-bit system. For an accurate approximation, the ``gaps'' on the upper boundary should be taken small compared to the remaining intervals.}
    \label{fig:discop}
\end{figure}

We define a ``discretization'' operator that maps from the Hilbert space of states for some holographic CFT to the Hilbert space describing the states of some collection of BCFT-bits. The operator is defined using the Euclidean path integral shown in figure \ref{fig:discop}. Explicitly, the matrix elements of the operator are defined as
\be\label{eq:discretization}
\langle \phi_{i}^0 | M_{H_\epsilon} |\phi^0 \rangle = \int_{\phi^- = \phi^0}^{\phi^+ = \phi_i^0} [d \phi] e^{-S_{BCFT}} \; ,
\ee
where $\phi^0$ is a field configuration in the original CFT, $\phi_i^0$ is a field configuration in the $i$th BCFT, and $H_\epsilon$ is a geometry with height $\epsilon$ in Euclidean time whose lower boundary is $M$ and whose upper boundary is $\{\tilde{M}_i\}$. A similar operator was discussed in \cite{Ohmori:2014eia} when discussing regularizations of entanglement entropy.

For any state $|\Psi_0 \rangle$ of the original CFT, we can then define an approximated state
\be
|\Psi \rangle = M_{H_\epsilon} |\Psi_0 \rangle \; .
\ee
In cases where the original state is produced by a Euclidean path integral with no operator insertions in the interval $[0, \epsilon]$ we can obtain a more accurate approximation to the original state by taking
\be
|\Psi \rangle = M_{H_\epsilon} e^{\epsilon H} |\Psi_0 \rangle \; .
\ee
This precisely reproduces the original construction of \cite{VanRaamsdonk:2018zws}. However, with or without the extra $e^{\epsilon H}$, the new state should give a good approximation to the original one in a limit where we take $\epsilon$ and the size of the added boundary components in $H_\epsilon$ to zero.

\section{Holographic duals of BCFT-bit states}\label{sec:holographicduals}

Adding boundary components to a holographic CFT has the effect of modifying the dual geometries to include ``end-of-the-world'' (ETW) branes that end on these boundary components. These are regions of the full spacetime that cap off the dual geometry in some way, which may involve a degeneration of the internal space in a higher-dimensional description of the geometry and/or some string theory branes.\footnote{See \cite{DHoker:2007zhm, DHoker:2007hhe, Aharony:2011yc,Assel:2011xz} for explicit microscopic examples.}

In this paper, we will make use of an effective description of these end-of-the-world branes, described in  \cite{Karch:2000gx,takayanagi2011holographic,fujita2011aspects}, where the ETW brane is modeled as a codimension-one spacetime boundary. We will focus on the simple case of $1+1$ dimensional boundary conformal field theories with central charge $c$ dual to $2+1$ dimensional AdS spacetimes with AdS length scale $\ell = (2c/3) G$. We take the tension of the brane to be $T / (8 \pi G)$, so that the full gravitational action is\footnote{More generally, there may be additional terms in the brane action, for example couplings to bulk scalars. However, this simple prescription already reproduces many features of BCFTs and will suffice for our purposes.}
\begin{align}
\label{action}
    I_{bulk} + I_{ETW} = \frac{1}{16\pi G}\int_N d^{2+1}x \sqrt{-g}(R - 2\Lambda) + \frac{1}{8\pi G} \int_Q d^{2}x \sqrt{-h}(K - T),
\end{align}
where $Q$ is the ETW brane, and terms at the AdS boundary are omitted.

This action (\ref{action}) leads to the usual Einstein's equations along with the boundary equation 
\begin{align}\label{eq:ADSBCFTbc}
    K_{ab} - K h_{ab}= -T h_{ab}.
\end{align}
As argued by Takayanagi \cite{Takayanagi:2011zk}, the tension parameter $T$ is positively related to the boundary entropy parameter $g_B$
as\footnote{The parameter $g_B$ may be defined as the regulated disk partition function for the BCFT or via the vacuum entanglement entropy for an interval of length $L$ containing the boundary as $S(L) = \frac{c}{6}\log \left( \frac{\ell}{\epsilon}\right) + \log g_B$.} 
\be
\log g_B = \frac{\ell}{4 G} {\rm arctanh}(T \ell) \; .
\ee
We can think of $\log g_B$ as an CFT boundary/edge analogue of the central charge giving a measure of the number of degrees of freedom associated with the CFT boundary.

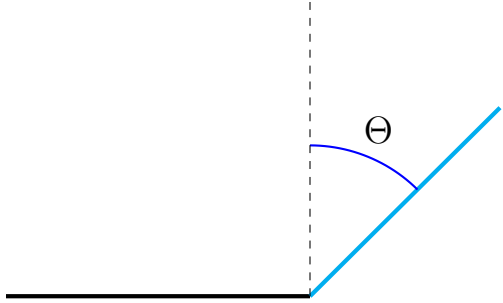
\begin{figure}
    \centering
    \begin{tikzpicture}
    
    \draw[ultra thick] (-4,0) -- (0,0);
    \draw[ultra thick,cyan] (0,0) -- (2.5,2.5);
    \draw[thick,blue,domain=90:45] plot ({2*cos(\x)},{2*sin(\x)}); 
    \draw[dashed] (0,0) -- (0,4);
    \node at (0.9,2.2) {\Large{$\Theta$}};
    
    \end{tikzpicture}
    \caption{Geometry dual to the vacuum state of a BCFT on a half line.}
    \label{fig:angle}
\end{figure}

As an example, for a 1+1 dimensional CFT on a half-space $x<0$, the dual geometry is Poincar\'e-AdS 
\be
ds^2 = \frac{\ell^2}{z^2}(dz^2 - dt^2 + dx^2)
\ee
with an ETW brane along an $AdS^2$ slice
\be
\frac{x}{z} = \tan \Theta
\ee
where the angle $\Theta$ is related to the tension parameter as
\be
 \sin \Theta = T \ell  \; .
\ee
This geometry is depicted in figure \ref{fig:angle}. 

\subsubsection*{Single gap geometry: Poincar\' e coordinates}

The simplest example of a map from CFT states to states of a disconnected BCFT system is the case where we introduce a single gap $[-\epsilon,\epsilon]$ in a 1+1 dimensional CFT, introducing a right boundary at $x = - \epsilon$ and a left boundary at $x = \epsilon$. 

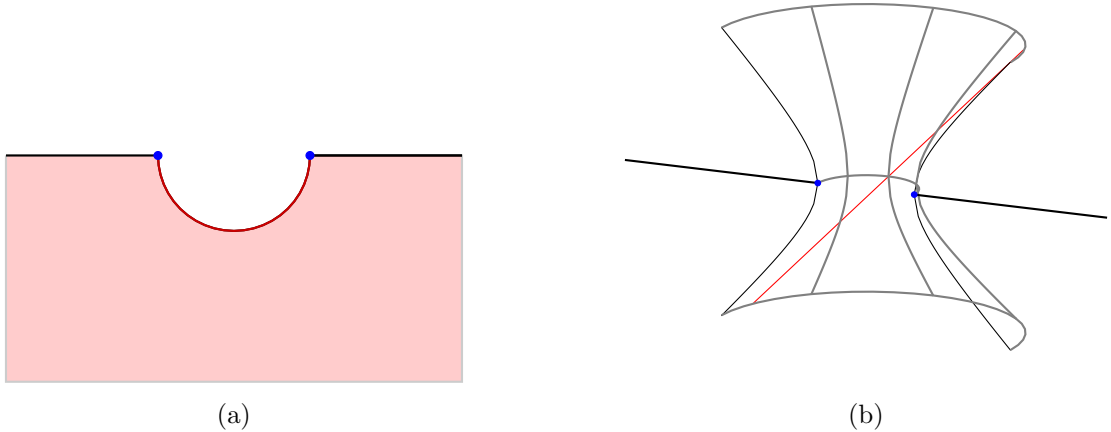
\begin{figure}
\begin{center}
\begin{subfigure}[b]{.45\textwidth}
\centering
\begin{tikzpicture}
    
    \draw[thick,variable=\x,domain=0:180] (-3,0) -- (-1,0) plot ({-cos(\x)},{-sin(\x)}) -- (3,0);
    \draw[thick,red,variable=\x,domain=0:180]  plot ({-cos(\x)},{-sin(\x)});
    \draw[thick,variable=\x,domain=0:180] (1,0)  -- (3,0);
    
    \draw[thick,variable=\x,domain=0:180,opacity=0.2,fill=red] (-3,0) -- (-1,0) plot ({-cos(\x)},{-sin(\x)}) -- (3,0) -- (3,-3) -- (-3,-3) -- (-3,0);
    
    \draw[blue] plot [mark=*, mark size=1.5] coordinates{(1,0)};
    \draw[blue] plot [mark=*, mark size=1.5] coordinates{(-1,0)};
     
\end{tikzpicture}
\caption{}
\label{fig:expandinga}
\end{subfigure}
\hfill
\begin{subfigure}[b]{.45\textwidth}
\begin{center}
    \tdplotsetmaincoords{15}{0}
    \begin{tikzpicture}[scale=0.7,tdplot_main_coords]
    \tdplotsetrotatedcoords{0}{25}{0}
    
    \begin{scope}[tdplot_rotated_coords]
    
    \draw[domain=-2.8:2.8,red] plot ({\x,\x,1});
    
    \draw[domain=1:3] plot ({\x,sqrt(\x^2-1),0});
    \draw[domain=1:3] plot ({\x,-sqrt(\x^2-1),0});
    
    \draw[domain=1:3] plot ({-\x,sqrt(\x^2-1),0});
    \draw[domain=1:3] plot ({-\x,-sqrt(\x^2-1),0});
    
    \begin{scope}[canvas is xz plane at y=0]
    \draw [gray,thick,domain=0:180] plot ({1*cos(\x)}, {1*sin(\x)});
    \end{scope}
    
    \begin{scope}[canvas is xz plane at y=2.82]
    \draw [gray,thick,domain=0:180] plot ({3*cos(\x)}, {3*sin(\x)});
    \end{scope}
    
    \begin{scope}[canvas is xz plane at y=-2.82]
    \draw [gray,thick,domain=0:180] plot ({3*cos(\x)}, {3*sin(\x)});
    \end{scope}
    
    \begin{scope}[canvas is yz plane at x=0]
    \draw [gray,thick,domain=1:3] plot ({sqrt(\x^2-1)}, {\x});
    \draw [gray,thick,domain=1:3] plot ({-sqrt(\x^2-1)}, {\x});
    \end{scope}
    
    \draw [gray,thick,domain=1:3] plot ({\x/1.41}, {sqrt(\x^2-1)},{\x/1.41});
    \draw [gray,thick,domain=1:3] plot ({\x/1.41}, {-sqrt(\x^2-1)},{\x/1.41});
    
    \draw [gray,thick,domain=1:3] plot ({-\x/1.41}, {sqrt(\x^2-1)},{\x/1.41});
    \draw [gray,thick,domain=1:3] plot ({-\x/1.41}, {-sqrt(\x^2-1)},{\x/1.41});
    
    \draw[thick] (-5,0,0) -- (-1,0,0);
    \draw[blue] plot [mark=*, mark size=1.5] coordinates{(-1,0,0)};
    
    \draw[thick] (5,0,0) -- (1,0,0);
    \draw[blue] plot [mark=*, mark size=1.5] coordinates{(1,0,0)};
    
    \end{scope}
    
    \end{tikzpicture}
\end{center}
\caption{}\label{fig:expandingb}
\end{subfigure}
\caption{(a) Euclidean path integral for an entangled state of two BCFTs approximating the vacuum state of the parent CFT. (b) Lorentzian gravity dual for this state. The ETW brane is anchored on the CFT boundaries (blue dots), and forms a hyperbola in the bulk. By coupling the CFT boundaries we hope to find a causally connected bulk geometry. The null curve running along the black hole horizon is shown in red.}
\label{fig:expanding}
\end{center}
\end{figure}

In order to approximate the original CFT vacuum state, we define a state of the pair of BCFTs using the Euclidean path-integral geometry shown in figure \ref{fig:expandinga}, where the boundary geometry in $\tilde{H}$ is the circle $x^2 + \tau^2 = \epsilon^2$. In this case, the full Euclidean path integral used for computing observables has the geometry of a plane with a disk removed. Rescaling coordinates as $(x,z,t) \to (x,z,t)/\epsilon$, the dual geometry is Poincar\'e-AdS with an ETW brane at \cite{Rozali:2019day}
\begin{align}
\label{EucBrane}
    x^2+\tau^2 +(z+\tan\Theta)^2=\sec^2\Theta.
\end{align}
For example, when $T \ell=\sin\Theta=0$, the ETW brane geometry is a half sphere in Poincar\'e coordinates.

The $\tau = 0$ slice of this Euclidean geometry provides the initial data for the Lorentzian geometry associated to our state. To determine a dual Lorentzian geometry, we must specify which Hamiltonian to evolve the state with. The simplest Lorentzian geometry is obtained by analytic continuation. 

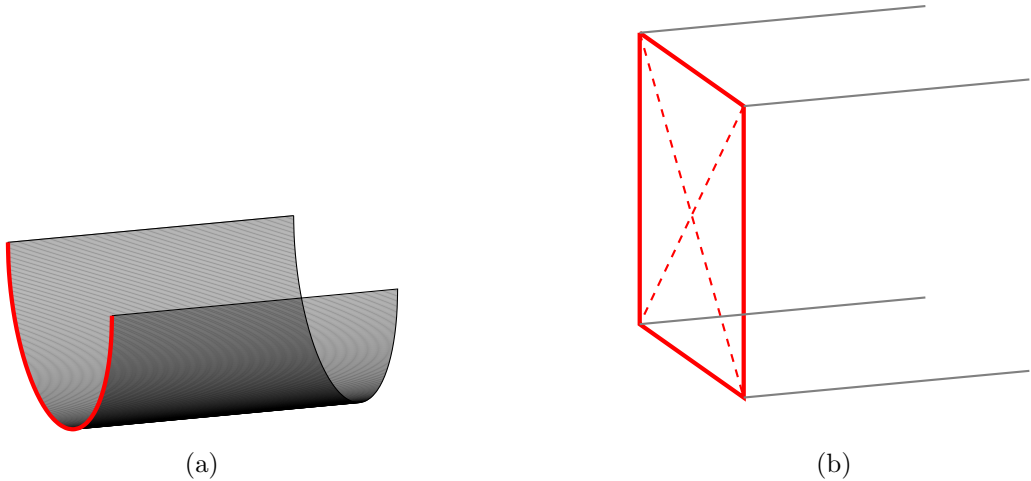
\begin{figure}
\begin{center}
\begin{subfigure}[b]{.45\textwidth}
\centering
\tdplotsetmaincoords{15}{0}
    \begin{tikzpicture}[scale=1,tdplot_main_coords]
    \tdplotsetrotatedcoords{0}{70}{0}
    \begin{scope}[tdplot_rotated_coords]
    
    \draw[domain=0:180,variable=\x,smooth] plot ({-2*cos(\x)}, {-2*sin(\x)},{4}) -- (2,0,0);
    \draw[domain=0:180,variable=\x,smooth] plot ({2*cos(\x)}, {-2*sin(\x)},{0}) -- (-2,0,4);
    
    \foreach \i in {1,...,360}
    {
        \draw[opacity=0.4,smooth] ({2*cos(\i/2)}, {-2*sin(\i/2)},{0}) -- ({2*cos(\i/2)}, {-2*sin(\i/2)},{4});
    }
    
    \draw[domain=0:180,variable=\x,smooth,ultra thick, red] plot ({2*cos(\x)}, {-2*sin(\x)},{0});
    
    \end{scope}
    \end{tikzpicture}
\caption{}
\label{fig:TFDpicsa}
\end{subfigure}
\hfill
\begin{subfigure}[b]{.45\textwidth}
\begin{center}
\tdplotsetmaincoords{15}{0}
    \begin{tikzpicture}[scale=1,tdplot_main_coords]
    \tdplotsetrotatedcoords{0}{70}{0}
    \begin{scope}[tdplot_rotated_coords]
    
    \draw[ultra thick,red] (-2,2) -- (2,2) -- (2,-2) -- (-2,-2) -- (-2,2);
    \draw[dashed,red,thick] (-2,2) -- (2,-2);
    \draw[dashed,red,thick] (2,2) -- (-2,-2);
    
    \draw[thick,gray] (2,2,0) -- (2,2,4);
    \draw[thick,gray] (-2,2,0) -- (-2,2,4);
    \draw[thick,gray] (-2,-2,0) -- (-2,-2,4);
    \draw[thick,gray] (2,-2,0) -- (2,-2,4);
    
    \end{scope}
    \end{tikzpicture}
\end{center}
\caption{}\label{fig:TFDpicsb}
\end{subfigure}
\caption{(a) The two-BCFT state in figure \ref{fig:expanding} approximating the vacuum of a single CFT can be understood as a thermofield double state of two BCFTs. (b) The dual geometry is a portion of the planar two-sided black hole capped off by an ETW brane.}
\label{fig:TFDpics}
\end{center}
\end{figure}

In this case, the ETW brane trajectory in the Lorentzian picture is a hyperbola
\begin{align}
    x^2 - t^2 +(z+\tan\Theta)^2=\sec^2\Theta.
\end{align}
The Lorentzian solution is shown in figure \ref{fig:expandingb}. In this conformal frame, the BCFT boundaries are dynamical, sitting on the two branches of the hyperbola $x^2 - t^2 = 1$. However, we can also change coordinates in the Euclidean case as
\be
\tau = \epsilon e^X \frac{\rho}{\sqrt{\rho^2+1}} \sin \phi  \qquad x = \epsilon e^X \frac{\rho}{\sqrt{\rho^2+1}} \cos \phi \qquad z = \epsilon \frac{e^X}{\sqrt{\rho^2 +1}}
\ee
so that hemispheres in Poincar\'e AdS map to constant $X$ slices of the Euclidean AdS cylinder with metric
\be
\label{cyl}
ds^2 = (\rho^2 + 1) d X^2 + \frac{d \rho^2}{(\rho^2 + 1)} + \rho^2 d \phi^2 \; .
\ee
In this case, the regions $x > \epsilon$ and $x < -\epsilon$ at $\tau = 0$ map to the regions $X>0$ for $\phi = 0, \pi$. In this description, the Euclidean path integral geometry defining the state of the pair of BCFTs is just the spatial geometry times the interval $\phi \in [0,\pi]$, and this gives exactly the thermofield double state. The Lorentzian geometry is a part of the planar two-sided BTZ black hole. In coordinates related to the Lorentzian Poincar\'e coordinates as
\be
t =  e^X \tan s \qquad x = e^X \sec s \sin w \qquad  z =  e^X \sec s \cos w\; ,
\ee
the metric becomes
\be
\label{swBH}
ds^2 = \frac{1}{\cos^2 w} (-ds^2 + dw^2 + \cos^2 s dX^2) 
\ee
the ETW brane trajectory is 
\be
\sinh X + \sec s \cos w \tan \Theta = 0 \; .
\ee
This description of the geometry is depicted in figure \ref{fig:TFDpics}.
The geometry (\ref{swBH}) has horizons at $s = \pm w$, so it is clear that we cannot traverse causally from one asymptotic region to the other, as required by the fact that our BCFTs are not interacting. 

The Lorentzian geometry we have described corresponds to evolving the $t=0$ state with a time-independent Hamiltonian in the conformal frame where the BCFTs are in thermal states; as we have seen, the BCFT edges are on accelerating trajectories in this frame. We could alternatively have chosen to consider a time-independent Hamiltonian in the original conformal frame so that the BCFT edges remained at fixed locations in the Poincar\'e description. In this case, the ETW brane geometry is more complicated to describe, but must also be non-traversable due to the expansion of the ETW brane.

\subsubsection*{Global BCFT-bit geometries}

\begin{figure}
    \centering
    \includegraphics[scale=0.4]{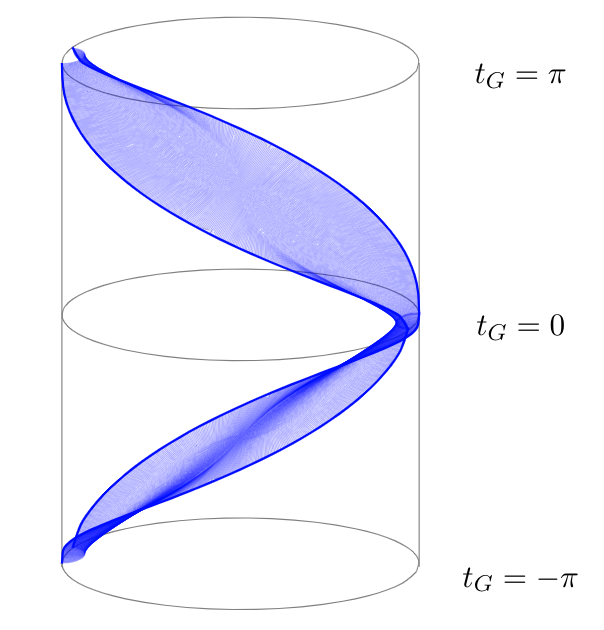}
    \caption{ETW brane in global Lorentzian AdS spacetime corresponding to the introduction of small disk-shaped hole in the Euclidean path integral constructing the vacuum state of a CFT on $S^1$. The $t_g=0$ slice of the remaining geometry contains an arbitrarily large part of the $t_G=0$ slice of the original global AdS spacetime.}
    \label{fig:globalbrane}
\end{figure}

Making use of our results for the simple case above, we can now understand the ETW brane trajectories in global AdS corresponding to a BCFT-bit state constructed from a path integral similar to the one in figure \ref{fig:review}.

First, we consider the geometry corresponding to the state of a BCFT constructed from the path-integral of figure \ref{fig:expandinga}, where we remove a small interval from the original CFT. To obtain the Lorentzian geometry in the tensionless case with $\Theta = 0$, we start from the Poincar\'e AdS description prior to the rescaling, where the ETW brane trajectory is 
\be
x^2 - t^2 + z^2 = \epsilon^2 \; .
\ee
The transformation
\be
z^{-1} = \sec r \cos t_G + \tan r \cos \theta \qquad t/z = \sec r \sin t_G \qquad x/z =\tan r  \sin \theta 
\ee
brings us to global AdS with metric
\be
ds^2 = \frac{1}{\cos^2 r} \left[ -dt_G^2 + dr^2 + \sin^2 r d \theta^2 \right]
\ee
where the ETW brane trajectory is
\be
\cos t_G  =  \frac{1 + \epsilon^2}{1 - \epsilon^2} \cos \theta \sin r \; .
\ee
We see that the ETW brane configuration is periodic in global time. In global coordinates, the size of the BCFT also oscillates with time. 

In the limit $\epsilon \to 0$, the ETW brane trajectory becomes the boundary of a Poincar\'e patch of global AdS.

\subsubsection*{General BCFT-bit geometries: Global}

\begin{figure}
    \centering
    \includegraphics[scale=0.70]{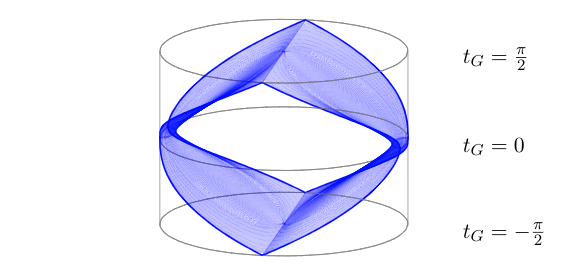}
    \caption{A global solution with two BCFT-bits. The state at $t_G=0$ is prepared by removing two disk shaped regions from the Euclidean path integral. The brane trajectories ensure there is no causal curve connecting the two BCFTs. The geometry to the future and past of the ETW branes is removed.}
    \label{fig:twobranes}
\end{figure}

Finally, we can consider multi-BCFT-bit states arising from path integrals similar to that in figure \ref{fig:review}. Here, the ETW brane can be described as the union of surfaces
\be
\cos t_G  =  \frac{1 + \epsilon^2}{1 - \epsilon^2} \cos (\theta - \theta_n) \sin r 
\ee
which correspond to the ETW branes associated with gaps of with $2 \epsilon$ centered at $\theta = \theta_n$. These are the surfaces of figure \ref{fig:globalbrane} shifted in the angular direction of the cylinder by angle $\theta_n$. Figure \ref{fig:twobranes} shows the resulting geometry for the case of two BCFT bits.

Since the local geometry is pure AdS and the individual ETW branes satisfy the equations of motion locally, the configurations with multiple ETW branes satisfy the gravitational equations up to the locations where the various ETW branes intersect. To understand the geometry here or at later times would require a more microscopic understanding of the ETW brane physics. 

In the limit $\epsilon \to 0$, the region under control becomes the intersection of Poincar\'e patches with cusps at $t_G = 0, \theta = \theta_n$. Taking $N$ boundaries with $\theta_n = 2 \pi n /N$ (where $\epsilon \ll 2 \pi/N$) and taking $N \to \infty$, this region approaches the full domain of dependence of the $t_G=0$ slice of our original AdS spacetime, as argued in \cite{VanRaamsdonk:2018zws}.

\subsection{Causal structure}

For these general BCFT-bit geometries, it must be the case that we cannot pass from one asymptotic region to another, since the BCFTs do not interact with one another. We have already argued for this directly for the two BCFT-bit geometry based on the description shown in figure \ref{fig:TFDpicsb}.

Alternatively, we can understand the non-traversability in Poincar\'e coordinates, noting that that the CFT boundary trajectories are the solutions to
\begin{align}
    x^2-t^2=1
\end{align}
and so asymptote the the $x=\pm t$ light rays at late and early times. Considering the full Poincar\'e boundary, there are no causal paths connecting the left and right regions. According to the results of \cite{gao2000theorems}, there cannot be any through the bulk either, since pure AdS satisfies the null curvature condition, and this ensures that no path through the bulk geometry can be faster than a path along the boundary. Finally, it follows immediately that there cannot be causal paths between the left and right asymptotic regions in our original geometry, since this is a subset of Poincar\'e AdS.

Since bulk causality is guaranteed to be consistent with boundary causality whenever the null energy condition holds \cite{gao2000theorems}, we can see that the CFT boundary trajectories along with the NEC guarantee non-traversability. In the following sections, we consider various ways to make the BCFTs interact that ultimately render the geometry traversable. In section \ref{sec:doubletrace} we add a double trace coupling between the BCFTs which leads to bulk violations of the NEC, while in section \ref{sec:intermediateCFT} we consider interactions that modify the boundary trajectories in the Poincar\'e description.

\section{Double-trace couplings of edge operators}\label{sec:doubletrace}

We have seen that the original BCFT-bit geometries can be understood as non-traversable multi-boundary wormholes. The non-traversability is implied by the fact that there are no interactions between the individual BCFTs. 

In this section, we investigate the effects of restoring interactions between the BCFTs in a minimal way, by adding in a coupling between low-dimension boundary operators on neighboring BCFTs boundaries. We would like to show that this renders the wormholes traversable, so that it is possible to move causally between asymptotic regions. 

Our construction is motivated by and technically similar to the discussion by Gao, Jafferis and Wall \cite{gao2017traversable}, who considered adding an interaction between two CFTs in the thermofield double state to make the dual two-sided black hole geometry traversable. The interaction was a time-dependent coupling between local operators of the two CFTs integrated over the spatial directions of the CFT.

In our case (focusing on the two-BCFT situation of  figure \ref{fig:expanding}), we start with two BCFTs in the thermofield double state. Our interaction is again a time-dependent coupling between local operators, but these are localized to the BCFT boundaries.

We have explicitly that 
\begin{align}
\label{LRcoupling}
    H = H_L + H_R + h(s) \psi_L(-s) \psi_R(s), 
\end{align}
where $\psi_L$ and $\psi_R$ are boundary operators for the left and right BCFTs. The coordinate $s$ represents the proper time along the boundaries, related to the Minkowski coordinates describing the Poincar\'e patch boundary according to\footnote{For technical convenience below, we are taking the variable $s$ to move backward in time on the left edge, but the coupling is between operators at the same Poincar\'e time.}
\begin{align}
    x = \pm\cosh(s), \nonumber \\
    t = \pm \sinh(s).
\end{align}
We would like to understand if this interaction between the left and right boundaries renders the bulk geometry traversable. In particular we will follow \cite{gao2017traversable} and choose the simple coupling, 
\begin{align}
    h(s) = h \theta(s).
\end{align}

Before adding this coupling, our geometry is that of a planar two-sided BTZ-black hole divided in half along the spatial direction, with one half removed and replaced by an ETW brane (figure \ref{fig:TFDpics}). As discussed in \cite{gao2017traversable}, the averaged null energy condition (ANEC) usually ensures that this geometry and perturbations to it are not traversable from one side to the other. But the double trace deformation of can lead to an ANEC-violating stress tensor, and a bulk null geodesic that travels from one asymptotic region of the black hole to the other. In the next subsection we find a sufficient condition for the existence of such geodesics in the perturbed geometry. In the following subsection, we show that the perturbations generated by our double-trace interaction satisfy this condition, so that the perturbed geometry is traversable.

\subsection{Stress tensor condition for traversability}\label{sec:STcondition}

Recall that the brane solution is, for $T=0$, 
\begin{align}
    x^2-t^2+z^2=1,
\end{align}
We will find it convenient to use light cone coordinates $v=x-t,u=x+t$. To look for geodesics that traverse this geometry, we will study null geodesics at $v=0$ and depth $z=z_0>1$. We can observe that at $v=0$, the brane is at depth $z=1$, so these do not collide with the brane. Further, all of these geodesics nearly traverse from one BCFT region to the other, in a sense we can make precise. 

Consider that at large $u$ we can deflect the geodesic, travelling initially at $v=0,z=z_0$, to travel slowly towards small $z$, so that $\dot z=-\sqrt{\epsilon}$. Then if we take $\dot x=1- \epsilon/2$, the geodesic remains null but reaches the boundary before travelling to $v=\sqrt{\epsilon}z_0/2$. Thus by incurring an arbitrarily small null delay we can travel from this geodesic to the boundary. Below, we will show that turning on the coupling \ref{LRcoupling} deflects these geodesics so that at large $u$ they reach $v(\infty)<0$. In the asymptotic region though the spacetime is locally pure AdS, so we can apply the above observation and bring these geodesics to the boundary, still at $v<0$. We can apply this reasoning twice to have the geodesic emerge from the left BCFT region, travel through the perturbed region and reach some $v<0$, then travel to the right BCFT region again. 

Given this reasoning, we find it suffices to study null geodesics at $v=0,z=z_0>1$. Consider a metric perturbation
\begin{align}
    ds^2 = \frac{\ell^2}{z^2}(dz^2+dudv+ \delta g_{\mu\nu} dx^\mu dx^\nu),
\end{align}
where we work in Fefferman-Graham gauge so that $\mu,\nu$ refer only to the light-cone coordinates. Working to first order in the metric perturbation, the $v$ component of the geodesic equation gives
\begin{align}
    \frac{d^2v}{du^2} =  -\Gamma^v_{uu} = -  \partial_u \delta g_{uu}.
\end{align}
Integrating with respect to $u$ twice, we obtain
\begin{align}
\label{vu}
    v(u) = -\int_{-\infty}^{u} du'\, \delta g_{uu}(u'),
\end{align}
where we chose $u$ to parameterize the brane. 

Next we should study Einstein's equations and try to relate this integral of $h_{uu}$ to the stress tensor. The $uu$ component of Einstein's equation is
\begin{align}
-\frac{z}{2}\partial_z\left( \frac{1}{z}\partial_z \delta g_{uu}\right) = 8\pi G T_{uu}.
\end{align}
Integrating this to find the metric perturbation in terms of $T_{uu}$ and inserting this expression into (\ref{vu}) we find that
\begin{align}\label{eq:traversabilitycondition}
   v(\infty) = - \int_{-\infty}^{+\infty} du\, \delta g_{uu} = 16 \pi G \int_{+\infty}^z dz' z' \int_{+\infty}^{z'} \frac{dz''}{ z''} \int_{-\infty}^{+\infty} du\, T_{uu}(u,v,z'') \; ,
\end{align}
where we have interchanged the order of integration. Our solution assumes $T_{uu}$ goes to zero at large $z$ sufficiently rapidly. The condition for traversability is $v(\infty) < 0$. Notice that the ANEC (holding for all constant $z$, $v=0$ null lines) is sufficient to give non-traversability, as expected.

\subsection{Correction to the stress tensor from point-splitting}\label{sec:stresstensorcorrection}

We would now like to calculate the bulk stress tensor that arises from a perturbation of the type (\ref{LRcoupling}) and see that it satisfies the condition $v(\infty) < 0$ for $v(\infty)$ given in (\ref{eq:traversabilitycondition}).

Consider a bulk scalar field $\varphi$ with mass $m$, dual to a CFT operator $\mathcal{O}_R$. We've included the subscript $R$ to indicate that $\mathcal{O}_R$ is an operator in the right BCFT. The bulk stress tensor for such a scalar field is given classically by
\begin{align}
    T_{\mu\nu} = \partial_\mu \varphi \partial_\nu \varphi - \frac{1}{2} g_{\mu\nu}g^{\rho\sigma} \partial_\rho\varphi \partial_\sigma \varphi - \frac{1}{2} M^2 \varphi^2.
\end{align}
Up to one loop order, this can be calculated using the point-splitting formula,
\begin{align}
    \langle T_{\mu\nu} (X) \rangle = \lim_{X\rightarrow X'} \left(\partial_\mu \partial_\nu' G(X,X') -\frac{1}{2} g_{\mu\nu}g^{\rho\sigma} \partial_\rho\partial_\sigma G(X,X') - \frac{1}{2}M^2 g_{\mu\nu}G(X,X') \right).
\end{align}
Our main technical challenge will be to calculate the two-point function $G(X,X')$ of the scalar field in the presence of the brane and with the boundary operator coupling turned on. 

Following \cite{gao2017traversable}, we begin with the observation
\begin{align}
    \langle \varphi^H(t)\varphi^H(t') \rangle = \langle U^{-1}(t,t_0) \varphi^I(t)U(t,t_0)U^{-1}(t',t_0)\varphi^I(t')U(t',t_0)\rangle \nonumber,
\end{align}
where $U$ is just the interaction part $U(t_0,t)=\mathcal{T}e^{-i\int_{t_0}^t dt \delta H}$, $\varphi^H$ has been time evolved in the Heisenberg picture (using the full Hamiltonian), and $\varphi^I$ is time evolved in the interaction picture sense (using only the unperturbed Hamiltonian). Next, we expand out the unitaries to first order,
\begin{align}
    \langle \varphi^H(t)\varphi^H(t') \rangle = \langle \varphi^I(t)\varphi^I(t') \rangle -i\int ds h(s) \Big(\langle [\psi_L(-s)\psi_R(s),\varphi(t)]\varphi(t') \rangle \nonumber \\
    +\langle \varphi(t) [\psi_L(-s)\psi_R(s),\varphi(t')] \rangle \Big). \nonumber 
\end{align}
Now we use large $N$ factorization and the fact that $\psi_L$ will commute with all the operators in the right wedge to find
\begin{align}
    \langle \varphi^H(t)\varphi^H(t') \rangle = \langle \varphi^I(t)\varphi^I(t') \rangle - i \int ds \,h(s) \Big( \langle \varphi(t') \psi_L(-s)\rangle \langle [\psi_R(s),\varphi(t)]\rangle \nonumber \\ 
    + \langle \varphi(t) \psi_L(-s) \rangle \langle [\psi_R(s),\varphi(t')]\rangle \Big)\nonumber
\end{align}
To calculate this first order correction term in the two point function we need the bulk-to-edge propagator $\langle \varphi(t) \psi_R(s)\rangle$. From this, we can use the KMS condition
\begin{align}
    \langle \varphi(t) \psi_L(s) \rangle = \langle \varphi(t) \psi_R(s+i\beta/2) \rangle
\end{align}
to learn the correlator $\langle \varphi(t) \psi_L(s)\rangle$, and we can easily construct the retarded propagator $\langle [\psi_R(s),\varphi(t')]\rangle$. To get the bulk-to-edge propagator, we begin with the bulk-to-boundary two-point function, which we calculate in the presence of a zero tension ETW brane in appendix \ref{sec:bulkboundarypropagator}. Then in appendix \ref{sec:edgepropagator} we use a boundary operator expansion to extract the bulk-to-edge propagator.

The final result for the bulk-to-edge propagator is
\begin{align}
    \langle \varphi(z,x,t) \psi(s) \rangle = \frac{2^\Delta z^\Delta}{[1+\tilde{r}^2-2x\cosh(s) + 2t\sinh(s)-i\epsilon]^{\Delta}}
\end{align}
The retarded propagator $\langle [\varphi(t),\psi_R(t')]\rangle$ is then
\begin{align}
    \langle [\varphi(t),\psi_R(s)]\rangle
    &= -2i\sin (\pi \Delta )\theta\left( -(1+\tilde{r}^2-2x \cosh s + 2t \sinh(s) )\right) |\langle \varphi(t)\psi_R(s) \rangle|
\end{align}
with the two-point function as given above.

We are interested in $\langle T_{uu}\rangle$. Using the point splitting formula, this is
\begin{align}\label{eq:tuulimit}
    \langle T_{uu}(u,v=0,z)\rangle = \lim_{u\rightarrow u'} \partial_u\partial_{u'}G(u,u').
\end{align}
It is straightforward to check that the contribution to $\langle T_{uu}(u,v=0,z)\rangle$ from the unperturbed two-point function is zero, so it is only the first order correction term we need to consider in the above.

\begin{figure}
    \centering
    \includegraphics[scale=0.5]{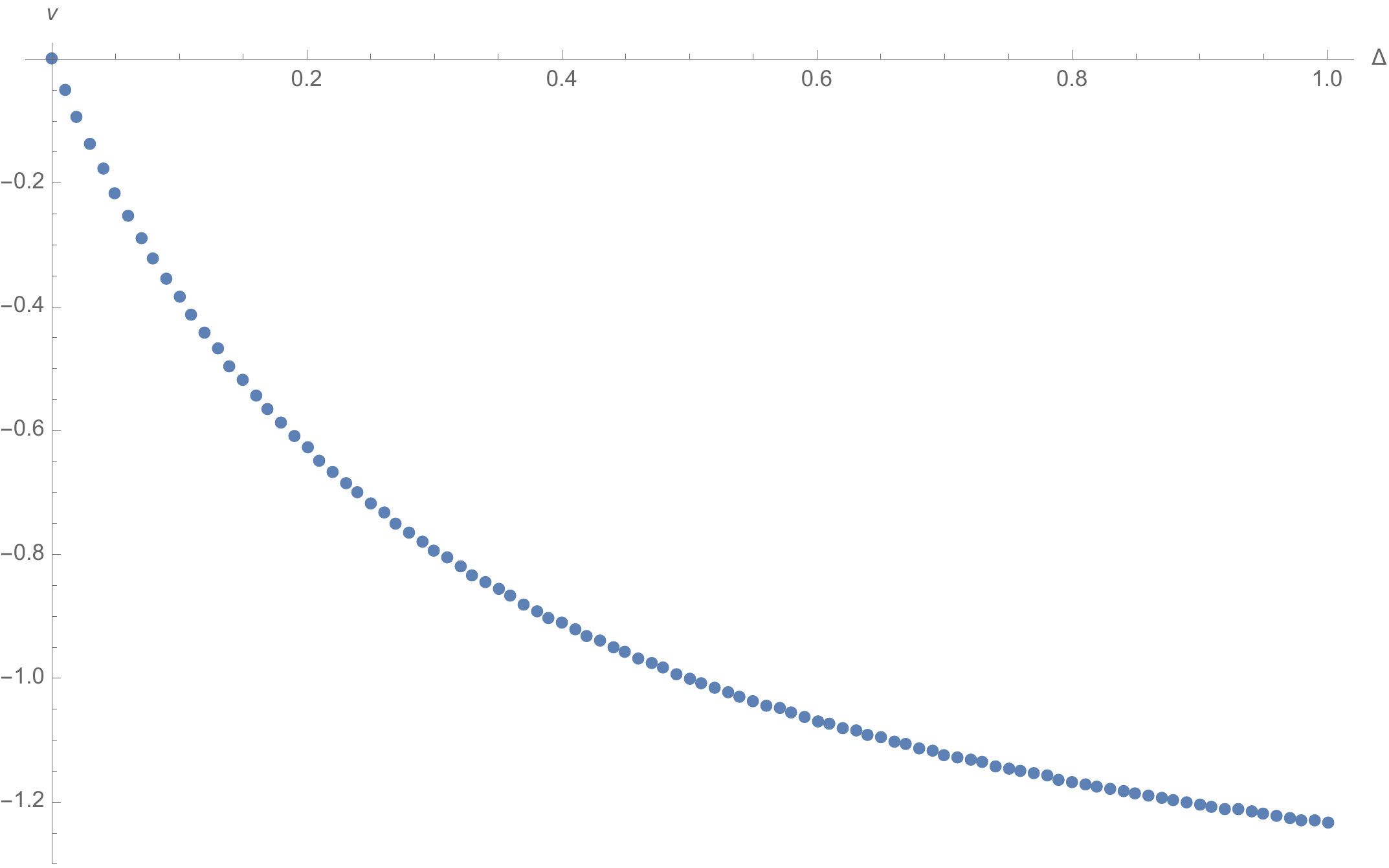}
    \caption{The deflection of the null ray at $v=0$ plotted as a function of the operator weight of the deforming operator $\psi$. Notice that the deflection is negative, indicating the geometry has become traversable. The deflection vanishes for $\Delta=0$, which corresponds to taking $\psi$ to be the identity. The vertical scale is measured in units of $1/16\pi G_N$.}
    \label{fig:deltavplot}
\end{figure}

It remains to insert the explicit forms of the propagators to find $G(u,u')$, calculate $T_{uu}$, and then insert this into the traversability condition (\ref{eq:traversabilitycondition}). We take this up in appendix \ref{sec:stresstensorcalc}. We find that the integrations factor,
\begin{align}
    v(\infty)=16\pi G_N h\left(\int_{+\infty}^{z} dz' z' \int_{+\infty}^{z'} \frac{dz''}{z''} F[z'',\Delta]\right)\left(\int_{-\infty}^{+\infty}d\tilde{u} \, H(\tilde{u},\Delta)\right).
\end{align}
where the functions $F[z,\Delta]$ and $H[\tilde{u},\Delta]$ are defined in appendix \ref{sec:stresstensorcalc}. These integrals may be done numerically for fixed values of $\Delta$. We plot $v(\infty)$ at $z=1$ in figure \ref{fig:deltavplot}, for $\Delta = i/100$, $i\in\{0,...,100\}$. We've chosen $h=1$ which leads to $v(\infty)<0$ for all $0\leq \Delta \leq 1$. Note that this also matches with the choice of sign for the coupling in \cite{gao2017traversable} used to render the two sided black hole traversable.

In summary, we find that by turning on an appropriate coupling between the two boundaries, the geometry becomes traversable from one asymptotic region to the other. In appendix \ref{sec:backreaction}, we also consider perturbations to the brane trajectory itself. We show that along the $v=0$, $z=1$ lightlike geodesic which lies inside the ETW brane in the unperturbed geometry, the perturbation to the brane trajectory moves toward smaller $z$ for large $u$, indicating that the ETW brane expands more slowly into the $z$ direction once the boundaries are coupled.

\section{Coupling via an intermediate CFT}\label{sec:intermediateCFT}

In the last section, we saw that adding a perturbative coupling between two BCFTs in the TFD state allows matter to travel from one BCFTs to the other. The metric perturbation is controlled by $G_N \langle T_{\mu\nu}\rangle$, and this comes in as an ${\cal O}(1/N^2)$ correction, so we could say that this is a geometry which is traversable at a quantum level. 

To make a classically traversable geometry, we could consider adding a large number (${\cal O}(N^2)$) of such couplings, as in the construction of \cite{maldacena2018eternal} for AdS$_{1+1}$ spacetimes. To achieve something like this for holographic BCFTs we can couple the two theories via some auxiliary degrees of freedom, following \cite{VanRaamsdonk:2020tlr}.\footnote{In a gauge theory, if the auxiliary degrees of freedom transform in a non-trivial representation of the gauge group of our original theories (e.g. bifundamental matter for the gauge groups associated with neighboring boundaries), then a coupling $\tr(M_L {\cal O}_{Aux})$ involving an adjoint operator $M_L$ on the left boundary and an operator ${\cal O}_{Aux}$ built from the original degrees of freedom can be understood as a coupling that involves ${\cal O}(N^2)$ separate matrix components of $M_L$.}

We take our auxiliary degrees of freedom to be some intermediate holographic CFT whose central charge we can control. In this case, the boundaries in the BCFT-bit picture become interfaces between our original bulk CFT and the new auxiliary CFT. 

If we take the auxiliary CFT to be the same as the original bulk CFT and the interface to be trivial, we have the original state of our original CFT. In the limit where the central charge of the auxiliary CFT goes to zero and the interface becomes the boundary used in our BCFTs, we get back the BCFT-bit state. So by varying the properties of the auxiliary CFT and interface, we will get a family of dual geometries that interpolate (in solution space) between the connected spacetime dual to the original CFT state and the multi-boundary wormhole state dual to the BCFT-bit state.

We will model these dual geometries using a bottom-up holographic setup (introduced in \cite{simidzija2020holo})  where the interface is a constant-tension interface brane. This separates a region of spacetime described by the low-energy effective theory associated with our original CFT from regions of spacetime described by a low-energy effective theory associated with the auxiliary CFT. 

For the new geometries corresponding to the coupled BCFTs, we would like to understand when we find traversability between asymptotic regions corresponding to the original CFT. In some cases, we may find that the new geometry is traversable, but only by causal paths that intersect the interface brane.  This would generally be impossible for an observer made from particles of the effective theory dual to our original CFT, since the regions of spacetime associated with the auxiliary degrees of freedom would in general have different light fields/particles. Thus, we will investigate the more refined question of when the region of spacetime described by the original effective theory becomes traversable. We will refer to this property as ``pleasant traversability.''

The specific states we will discuss are those created by the Euclidean path integrals in figure \ref{fig:perturbedPI}. To find the dual geometry in each case, we consider the Euclidean geometry associated with the doubled bra-ket path integral. This is time-symmetric, and the geometry of the slice left fixed under Euclidean time-reversal (together with the condition that time-derivatives vanish) gives the initial data for our Lorentzian evolution.

We emphasize that given the choice of state / initial data, we still have a choice of which CFT Hamiltonian to evolve the state with. On the gravity side, this corresponds to our choice of boundary conditions for the Lorentzian evolution. The simplest choice is the time-independent Hamiltonian where the interfaces are at fixed location so that the interval on which the auxiliary degrees of freedom live is of a fixed size. Another natural choice corresponds to simply taking the analytic continuation of the Euclidean geometry. In general, this corresponds to a time-dependent interface CFT Hamiltonian where the locations of the interfaces change with time in a way determined by the analytic continuation of their Euclidean trajectories. For our investigations, we will take this latter approach when finding explicit geometries, since it will allow analytic solutions, but we will also discuss the physics of the case with a time-independent CFT Hamiltonian.

\subsection{Holographic ICFT preliminaries}\label{sec:ICFTpreliminaries}

From the field theory perspective, the parameters that we use to characterize the interface theory are the central charges of the original and auxiliary CFTs, as well as the interface entropy $\log g_I$. The interface entropy appears as a term in the vacuum entanglement entropy for an interval of size $2L$ centered on the interface:
\begin{align}\label{eq:defectentropy}
    S(2L) = \frac{c_1}{6} \log \frac{L}{\epsilon} + \frac{c_2}{6} \log \frac{L}{\epsilon} + \log g_{I} \; .
\end{align}
The interface entropy can be understood as a measure of the number of degrees of freedom associated with the interface.

To study the gravity dual of holographic interface CFTs, we will follow\footnote{See also \cite{Bachas:2020yxv} for recent discussions of interface theories in holography.} \cite{Erdmenger:2014xya,simidzija2020holo} and employ a bottom-up model in which the CFT interface extends into the bulk as a dynamical interface brane separating regions with different low-energy effective theories. We take the bulk gravity theories associated with the original and auxiliary CFTs to be Einstein gravity with AdS lengths $\ell_i$ related to the central charge parameters,
\begin{align}
    c_{1,2} = \frac{3\ell_{1,2}}{2G_N}.
\end{align}
We take the interface brane to have a constant tension and model its dynamics using the Israel junction conditions, which require in this case that the induced metric on the brane is the same when using the bulk metric on either side, and that the extrinsic curvatures vary across the brane as
\begin{align}
    K_{ab}^1- K_{ab}^2 = T h_{ab} \; ,
\end{align}
where $K_{ab}^i$ is the extrinsic curvature of the brane, computed using either the metric $g^1_{\mu\nu}$ or $g^2_{\mu\nu}$. 

As discussed in \cite{simidzija2020holo}, by finding the bulk solution corresponding to a planar interface and using the Ryu-Takayanagi formula to compute the entanglement entropy, we find that the tension parameter $T$ is related to the CFT parameters as
\begin{align}\label{eq:interfaceentropyandtension}
    \log g_I = \frac{\ell_1+\ell_2}{4G_N}\arctanh\left( \frac{T}{\frac{1}{\ell_1}+\frac{1}{\ell_2}} \right)+\frac{\ell_1-\ell_2}{4G_N}\arctanh\left( \frac{\frac{1}{\ell_1}-\frac{1}{\ell_2}}{T} \right)
\end{align}
We can observe that $T$ in the domain
\begin{align}\label{eq:InterfaceTdomain}
    \left|\frac{1}{\ell_1} - \frac{1}{\ell_2}\right| \leq T \leq \frac{1}{\ell_1} + \frac{1}{\ell_2}
\end{align}
maps to all real values of $\log g_I$. This motivates restricting $T$ to the above domain. 

\vspace{0.25cm}
\noindent \textbf{Circular interface solutions}
\vspace{0.25cm}
\label{sec:disk}

Let us first understand the geometries dual to states constructed via the Euclidean path integral shown in figure \ref{fig:perturbedPIb}, where we simply fill in the missing disc in the original path integral of figure \ref{fig:perturbedPIa} with an auxiliary CFT. 

The Euclidean path-integral used for computing observables in this state is the ``doubled'' path integral where the interface is now a circle in the full plane. This is associated with a dual gravity solution with an interface brane ending on a circle. From this, we will analytically continue to obtain a Lorentzian solution. In the Poincar\'e description, this corresponds to an evolution where the interface locations follow uniformly accelerating trajectories so that the interval size inflates. The energy density in the original CFTs away from the interface vanishes. Alternatively we can go to a conformal frame related to the one in figure \ref{fig:TFDpics}. Here, we have two copies of the thermofield double state of a theory on $\mathbb{R}$ with a single interface at $y=0$.  

To obtain the solution, consider first the interface CFT on a Euclidean cylinder with our original CFT (call this $CFT_1$) on the $y > 0$ region and the auxiliary CFT (call it $CFT_2$) on the $y < 0$ region, where $y$ is the coordinate along the length of the cylinder.

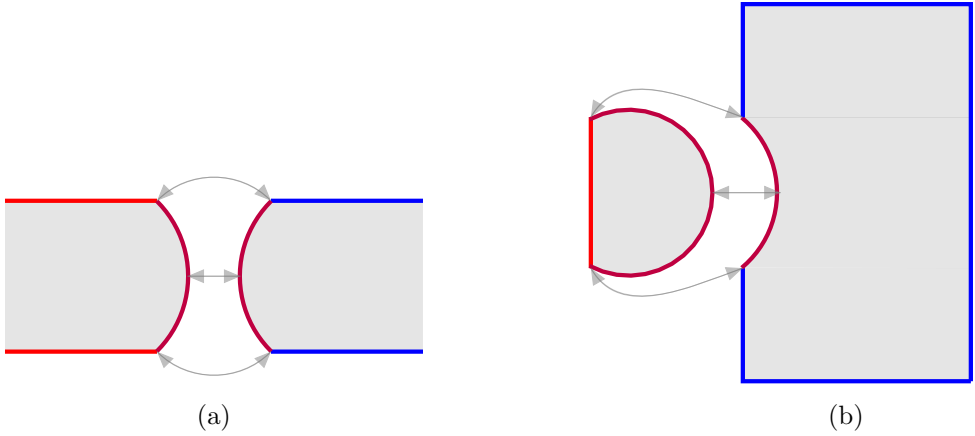
\begin{figure}
\begin{center}
\begin{subfigure}[b]{.45\textwidth}
\centering
\begin{tikzpicture}
    
    \draw[domain=45:-45,fill=gray,opacity=0.2] (-2,2) -- (0,2) -- plot ({1.41*cos(\x)-1},{(1.41*sin(\x))+1}) -- (0,0) -- (-2,0);
    
    \draw[ultra thick,red] (-2,2) -- (0,2);
    \draw[ultra thick,red] (-2,0) -- (0,0);
    
    \draw[ultra thick, purple,domain=-45:45] plot ({1.41*cos(\x)-1},{(1.41*sin(\x))+1});
    
    \draw[domain=135:225,fill=gray,opacity=0.2] (3.5,2) -- (1.5,2) -- plot ({1.41*cos(\x)+2.5},{(1.41*sin(\x))+1}) -- (1.5,0) -- (3.5,0);
    
    \draw[ultra thick,blue] (1.5,2) -- (3.5,2);
    \draw[ultra thick,blue] (1.5,0) -- (3.5,0);
    
    \draw[ultra thick, purple,domain=135:225] plot ({1.41*cos(\x)+2.5},{(1.41*sin(\x))+1});
    
    \draw[gray, -triangle 45,opacity=0.5] (0,2) to  [out=45,in=135] (1.5,2);
    \draw[gray, -triangle 45,opacity=0.5] (1.5,2) to  [out=135,in=45] (0,2);
    
    \draw[gray, -triangle 45,opacity=0.5] (0,0) to  [out=-45,in=-135] (1.5,0);
    \draw[gray, -triangle 45,opacity=0.5] (1.5,0) to  [out=-135,in=-45] (0,0);
    
    \draw[gray, -triangle 45,opacity=0.5] (0.4,1) -- (1.1,1);
    \draw[gray, -triangle 45,opacity=0.5] (1.1,1) -- (0.4,1);
     
\end{tikzpicture}
\caption{}
\label{fig:globalcoords}
\end{subfigure}
\hfill
\begin{subfigure}[b]{.45\textwidth}
    \begin{tikzpicture}[scale=1]
    
    \draw[fill=gray,opacity=0.2,domain=118:-118] (-1,-1) -- (-1,1) plot ({1.1*cos(\x)-0.5},{1.1*sin(\x)});
    
    \draw[ultra thick, red] (-1,-1) -- (-1,1);
    \draw[ultra thick, purple, domain=-118:118] plot ({1.1*cos(\x)-0.5},{1.1*sin(\x)});
    
    \fill[gray,domain=-50:50,opacity=0.2] (1,1) -- (1,2.5) -- (4,2.5) -- (4,1) --(1,1);
    \fill[gray,domain=-50:50,opacity=0.2] (1,-1) -- (1,-2.5) -- (4,-2.5) -- (4,-1) --(1,-1);
    
    \fill[opacity=0.2,domain=-50:50,gray] plot ({1+1.3*cos(\x)-0.85},{1.3*sin(\x)}) -- (1,1) -- (4,1) -- (4,-1);
    
    \draw[ultra thick, blue] (4,-2.5) -- (1,-2.5) -- (1,-1);
    
    \draw[ultra thick, purple, domain=-50:50] plot ({1+1.3*cos(\x)-0.85},{1.3*sin(\x)});
    
    \draw[ultra thick, blue] (1,1) -- (1,2.5) -- (4,2.5) -- (4,-2.5);
    
    \draw[-triangle 45,gray,opacity=0.5] (0.6,0) -> (1.5,0);
    \draw[-triangle 45,gray,opacity=0.5] (1.5,0) -> (0.6,0);
    
    \draw[-triangle 45, gray,opacity=0.5] (1,1) to  [out=160,in=60] (-1,1);
    \draw[-triangle 45, gray,opacity=0.5] (-1,1) to  [out=60,in=160] (1,1);
    
    \draw[-triangle 45, gray,opacity=0.5] (1,-1) to  [out=-160,in=-60] (-1,-1);
    \draw[-triangle 45, gray,opacity=0.5] (-1,-1) to  [out=-60,in=-160] (1,-1);
    
    \end{tikzpicture}
\caption{}\label{fig:Poincarecoords}
\end{subfigure}
\caption{Cut-away views of interface brane solutions in $AdS^3$ (a) In global coordinates, the interface is at $y=0$. The CFT interface extends into the bulk as an interface brane (purple). The two faces of the interface brane are described in (\ref{eq:y1}) and (\ref{eq:y2}). (b) In Poincar\'e coordinates, the CFT interface is a circle. The two faces of the interface brane form portions of spheres, see equation (\ref{eq:u2sol}).}
\label{fig:twodescriptions}
\end{center}
\end{figure}

On the gravity side the regions $\mathcal{M}_1$ and $\mathcal{M}_2$ on either side of the interface brane can be described using $AdS^3$ global coordinates 
\begin{align}
    ds^2_i = f_i(\rho) dy^2 + \frac{d\rho^2}{f_i(\rho)} +\rho^2 d\phi^2 \,\,\,\,\,\,f_i(\rho) = 1 + \lambda_i \rho^2.
\end{align}
We have introduced the parameter $\lambda_i=1/\ell_i^2$. These global coordinates are related to (radial) Poincar\'e coordinates by the transformation
\begin{align}\label{eq:globaltoradial}
    \sqrt{\lambda_i} y &= \ln r = \ln \sqrt{x^2+\tau^2+z^2}, \nonumber \\
    \sqrt{\lambda_i} \rho &= \tan\theta = \frac{\sqrt{x^2+\tau^2}}{z} ;\ ,
\end{align}
which maps $y = - \infty$ to the origin on the boundary and the interface on the boundary of the cylinder to the unit circle on the $z=0$ boundary.
 These two different descriptions are illustrated in figure \ref{fig:twodescriptions}. 

The necessary interface brane solutions appear already in \cite{simidzija2020holo}. The interface trajectories in the two components of space are described as\footnote{See eq. 4.11 of \cite{simidzija2020holo}. While they consider two interfaces ending at $y_2=\pm S/2$ for some parameter $S$, we have only one interface which ends at $y_2=0$.}
\begin{align}\label{eq:y1}
    \tanh(\sqrt{\lambda_1} y_1) = \frac{(T^2-\lambda_2+\lambda_1)}{2T\sqrt{\lambda_1}\sqrt{1+A\rho^2}}
\end{align}
\begin{align}\label{eq:y2}
    \tanh(\sqrt{\lambda_2} y_2) = \frac{(-T^2-\lambda_2+\lambda_1)}{2T\sqrt{\lambda_2}\sqrt{1+A\rho^2}}
\end{align}
with 
\begin{align}
    A = \lambda_1 - \frac{(T^2 - \lambda_2+\lambda_1)^2}{4T^2}.
\end{align}
Notice that we only have $y_1,y_2\rightarrow 0$ as $\rho\rightarrow \infty$ if $A>0$, which is indeed the case for $T$ in the allowed range (\ref{eq:InterfaceTdomain}). 

We would like to view these solutions in Poincar\'e coordinates. Making the transformation (\ref{eq:globaltoradial}) to Poincar\'e coordinates, we find that the description of the interface in each region corresponds to the $z>0$ part of the spherical surface
\begin{align}\label{eq:u2sol}
    x^2+\tau^2 + (z - \gamma_i)^2  = 1+ \gamma_i^2
\end{align}
where $\gamma_i$ is fixed in terms of $T, \lambda_1$, and $\lambda_2$ as 
\begin{align}
    \gamma_1 =  \frac{(T^2 - \lambda_2 + \lambda_1)}{\sqrt{4 \lambda_1 \lambda_2 - (T^2 - \lambda_1 - \lambda_2)^2 }} \,\,\,\,\,,\,\,\,\,\,\,\, \gamma_2 =  -\gamma_1 \; .
    \label{defGamma}
\end{align}

Considering the outer surface and comparing with (\ref{EucBrane}), we see that the brane trajectories for this interface case take the same form as the ETW-brane trajectories in the BCFT case with $T_{ETW}$ determined by $ T_{ETW}/\sqrt{\lambda_1^2 - T_{ETW}^2} = \gamma_1$. Thus, the geometries associated with states produced by the Euclidean path integral with a circular interface are not pleasantly traversable since the BCFT geometries not traversable. In fact, the geometries are not traversable at all from left to right, as we will argue more directly below.

\subsubsection*{Evolution with a time-independent Hamiltonian}

The non-traversability we have observed is related to the fact that in Poincar\'e coordinates, the interface trajectories in the Lorentzian picture follow the hyperbolic trajectories
\be
x^2 - t^2 = 1 \; .
\ee
On the other hand, we could evolve the same initial state with an interface CFT Hamiltonian that is static, so that the interfaces remain at $x = \pm 1$. In this case, we would still have initially zero energy density in the left and right bulk CFTs, but the interface and the interior CFT carry energy relative to their vacuum configurations. As a function of time, this energy should leak out into the exterior CFTs and move off to $x = \pm \infty$. After a long time, we expect the degrees of freedom in the vicinity of the interface to settle down to the vacuum configuration. In the next section, we'll understand the geometries dual to this vacuum configuration. 

\subsection{Strip geometries}
\label{sec:strip}

To construct the vacuum state of the interface CFT, we can consider the path integral of figure \ref{fig:perturbedPId} that is translation-invariant in the Euclidean time direction. The time-translation invariance in the Euclidean time direction means that the associated Lorentzian geometry will be static. The relevant geometries can be extracted from the results of  \cite{simidzija2020holo}, which considered the case of a strip of $CFT_2$ in the Euclidean time interval $[-S/2,S/2]$ of an infinite cylinder with $CFT_1$ living on the $|\tau| > S/2$ regions. The strip geometry we are interested in is obtained by taking the limit where $S$ is fixed and the radius of the cylinder goes to infinity, or equivalently, where $S \to 0$ with a fixed cylinder radius (section 4.5 of \cite{simidzija2020holo}). 

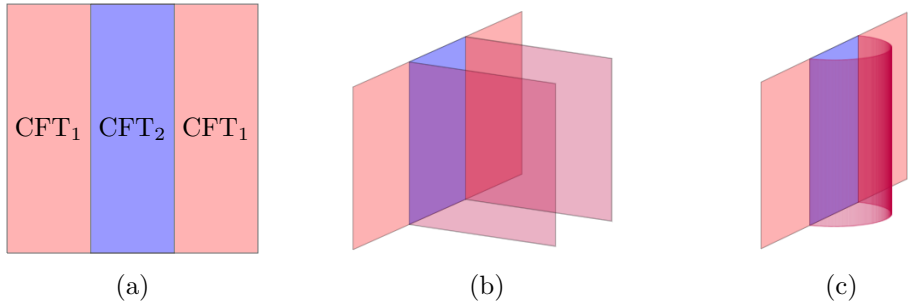
\begin{figure}
    \centering
\begin{subfigure}[b]{.3\textwidth}  
\centering
    \begin{tikzpicture}[scale = 0.55]
    
    \draw[gray] (-3,-3) -- (-3,3) -- (3,3) -- (3,-3) -- (-3,-3);
    
    \draw[gray,fill=red,opacity=0.3] (-3,-3) -- (-3,3) -- (-1,3) -- (-1,-3) -- (-3,-3);
    \draw[gray,fill=red,opacity=0.3] (3,-3) -- (3,3) -- (1,3) -- (1,-3) -- (3,-3);
    
    \draw[fill=blue,opacity=0.4] (-1,3) -- (1,3) -- (1,-3) -- (-1,-3) -- (-1,3);
    
    \node at (0,0) {\small{CFT$_2$}};
    \node at (2,0) {\small{CFT$_1$}};
    \node at (-2,0) {\small{CFT$_1$}};
    
    \end{tikzpicture}
     \caption{}
  \label{fig:phase0}
\end{subfigure}    
\begin{subfigure}[b]{.3\textwidth}
  \centering
  \includegraphics[width=.8\linewidth]{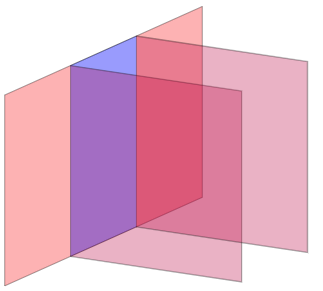}  
  \caption{}
  \label{fig:phase1}
\end{subfigure}
\begin{subfigure}[b]{.3\textwidth}
  \centering
  \includegraphics[width=.5\linewidth]{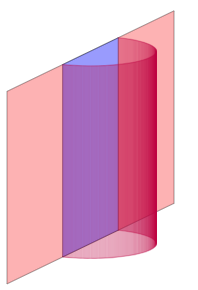}  
  \caption{}
  \label{fig:phase2}
\end{subfigure}
\caption{(a) Euclidean path integral geometry associated with the vacuum state of the interface CFT. (b),(c) Two possible topologies for the interface brane in the dual geometry. The configuration (b) is traversable from left to right but this requires crossing the interface brane. The configuration (c) is pleasantly traversable.}
\label{fig:strip-bothphases}
\end{figure}

At least for some ranges of parameters\footnote{See figure 10 of \cite{simidzija2020holo}, where $\kappa$ corresponds to our parameter $T$. The regions of the phase diagram labeled as ``bubble inside horizon'' and ``bubble outside horizon'' have a connected interface brane.} (when the central charge of the auxiliary CFT is greater than 1/3 of the central charge of the original CFT and the interface entropy is sufficiently high), this static geometry has a connected interface brane as shown in figure \ref{fig:phase2}, so it is possible to pass between the asymptotic regions associated with the left and right bulk CFTs without crossing the interface brane. In the other case, the geometry is still traversable, but not without passing through the interface brane. 

We can understand the distinction between the two cases here in terms of RG flow. The theory with two interfaces flows in the IR to a CFT (our original bulk CFT) with a single defect. In some cases, this defect is trivial; this corresponds to the case of a connected interface brane where the dual geometry is pleasantly traversable.

As we have argued above, these vacuum geometries should control the nature of traversability for any situation with a time-independent Hamiltonian where out CFTs have zero energy density asymptotically so that the degrees of freedom in the vicinity of the interface can settle to their vacuum configuration.

\subsection{Deforming the circle}

We have now understood the dual geometries associated with the Euclidean path integrals where the auxiliary theory has the geometry of a disk or a strip. In the case where we take the Lorentzian geometry to be simply the analytic continuation of the corresponding Euclidean one (which corresponds to a time-dependent interface in the disk case), the dual geometry is non-traversable in the disk case and traversable (sometimes pleasantly) in the strip case. 

In this section, we consider an intermediate situation where we deform the disk configuration perturbatively into an ellipse with the longer axis in the Euclidean time direction as shown in figure \ref{fig:perturbedPIc}. Continuing this deformation while keeping the width of the ellipse fixed, we would eventually arrive at the strip case, so the dual geometries we find should be intermediate between the non-traversable geometries considered in section \ref{sec:ICFTpreliminaries} and the traversable ones considered in section \ref{sec:strip}. In fact, we will see that traversability is restored even for a small deformation of the interface shape.

To motivate this, note that for path-integral state produced by a circular Euclidean interface, the CFT interface trajectories in the Poinacar\'e coordinates are hyperbolic trajectories ($x^2-t^2 = 1$) which asymptote to the light rays which pass through the origin. Thus, left and right regions housing the original bulk CFT degrees of freedom are spacelike separated. It is therefore impossible to travel through the bulk from one asymptotic region to the other even if we allow passing through the interface brane. To remedy this, we need to deform the Euclidean path integral so that the edge trajectories are changed. 

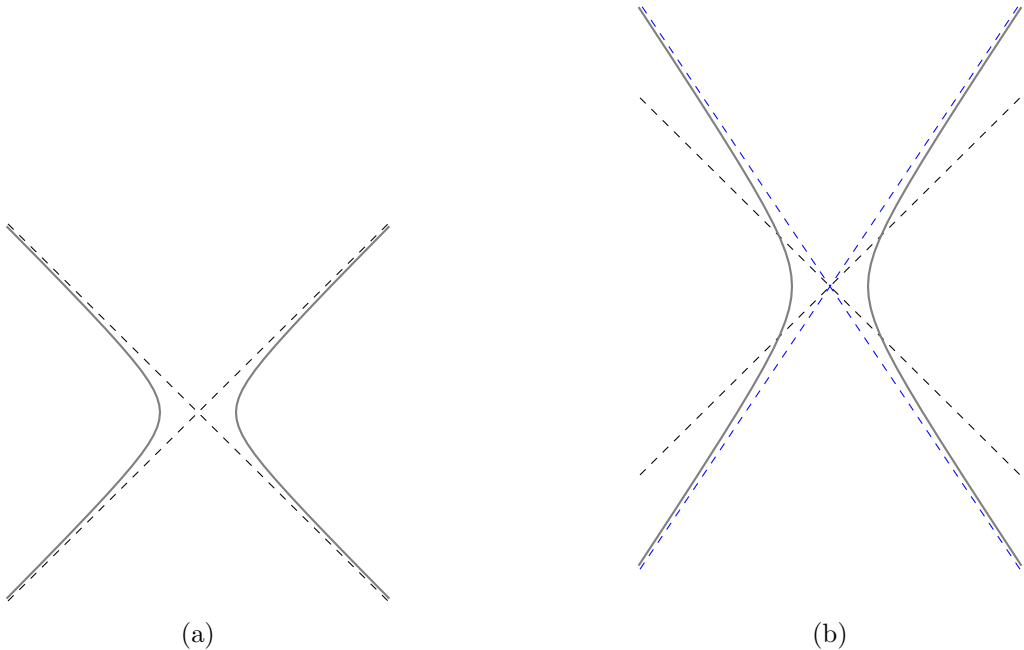
\begin{figure}
\begin{center}
\begin{subfigure}[b]{.45\textwidth}
\centering
\begin{tikzpicture}[scale=0.5]

\draw[dashed] (-5,-5) -- (5,5);
\draw[dashed] (-5,5) -- (5,-5);

\draw [gray,thick,domain=-2.3:2.3] plot ({cosh(\x)}, {sinh(\x)});
\draw [gray,thick,domain=-2.3:2.3] plot ({-cosh(\x)}, {sinh(\x)});

\end{tikzpicture}
\caption{}
\label{fig:unperturbededge}
\end{subfigure}
\hfill
\begin{subfigure}[b]{.45\textwidth}
\begin{center}
\begin{tikzpicture}[scale=0.5]

\draw[dashed] (-5,-5) -- (5,5);
\draw[dashed] (-5,5) -- (5,-5);

\draw[dashed,blue] (-5,-5*1.5) -- (5,5*1.5);
\draw[dashed,blue] (5,-5*1.5) -- (-5,5*1.5);

\draw [gray,thick,domain=-2.3:2.3] plot ({cosh(\x)}, {1.5*sinh(\x)});
\draw [gray,thick,domain=-2.3:2.3] plot ({-cosh(\x)}, {1.5*sinh(\x)});

\end{tikzpicture}
\end{center}
\caption{}\label{fig:perturbededge}
\end{subfigure}
\caption{(a) The Wick rotation of a circle, giving the edge trajectories when the path integral has a disk of CFT1. The edge trajectories asymptote to the light rays passing through the origin. The left CFT is behind a horizon from the perspective of the right CFT. (b) The Wick rotation of an ellipse which has been stretched in the $\tau$ direction. The edge trajectories now asymptote to timelike lines. The left and right CFTs are now causally connected via the intermediate CFT1 region.}
\label{fig:edgetrajectories}
\end{center}
\end{figure}

We'll deform from a circular interface to an elliptical one, with the ellipse elongated in the Euclidean time direction,
\begin{align}
    x^2 + \tau^2/b^2=1, \,\,\,\,\,\, b>1
\end{align}
The Wick rotated edge trajectories are shown in figure \ref{fig:edgetrajectories}. The edge trajectories now asymptote to $x = \pm t/b$ with $b>1$. Importantly, there is no horizon separating the left and right CFTs, so the bulk geometry is traversable, though it is not clear if it is pleasantly traversable. 

To study if the bulk geometry is pleasantly traversable we will need to look at the perturbed brane solution. In the next section we will establish that the bulk solutions have the following two features, 
\begin{enumerate*}
    \item Defining $|x|_{max}(t)$ as the maximal value of $|x|$ obtained by the interface brane along a constant time slice, we have that there exists $\epsilon>0$ such that $dx_{max}/dt \leq1-\epsilon$ for all $t$.
    \item Defining $z_{max}(t)$ as the maximal $z$ value obtained by the brane within the time interval $[-t,t]$, we have that $z_{max}<\infty$ for $t<\infty$.
\end{enumerate*}
These two features together suffice to render the bulk geometry pleasantly traversable, as we will establish below. Notice that the circular interface case fails to satisfy the first condition. 

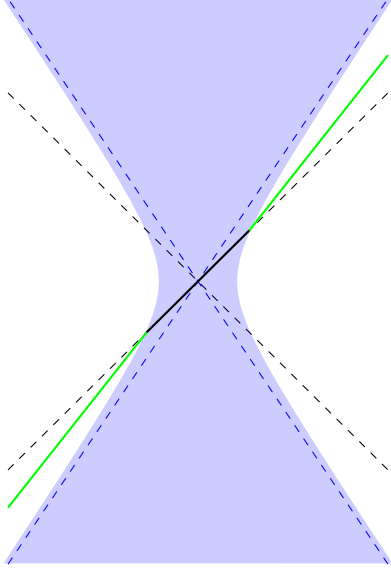
\begin{figure}
\begin{center}
\begin{tikzpicture}[scale=0.5]

\draw[dashed] (-5,-5) -- (5,5);
\draw[dashed] (-5,5) -- (5,-5);

\draw[dashed,blue] (-5,-5*1.5) -- (5,5*1.5);
\draw[dashed,blue] (5,-5*1.5) -- (-5,5*1.5);


\draw [blue,thick,domain=-2.31:2.31,fill=blue,opacity=0.2] plot ({cosh(\x)}, {1.5*sinh(\x)}) -- plot ({-cosh(\x)}, {-1.5*sinh(\x)});

\draw[thick] (-1.35,-1.35) -- (1.35,1.35);

\draw[thick,green] (-5,-6) -- (-1.35,-1.35);
\draw[thick,green] (1.35,1.35) -- (5,6);

\end{tikzpicture}
\caption{Construction of a causal curve that traverses from the left to the right CFT. The blue shaded region represents the projection of the brane onto the boundary, while the green and black segments are the projections of curves $\gamma_I$, $\gamma_{II}$ and $\gamma_{III}$. The green segment begins at an early time at the boundary, and extends slowly into the bulk. The black segment is at fixed $z=z_*$. $z_*$ is chosen as described in the main text to be outside the brane. The green segment should be adjusted to attach smoothly onto the black one. The second green segment extends slowly towards the boundary.}
\label{fig:edgetrajectories2}
\end{center}
\end{figure}

To see that conditions 1) and 2) together ensure pleasant traversability, we construct a causal curve passing from the left to the right CFT using only these assumptions. We construct this causal curve in three segments, which we show in figure \ref{fig:edgetrajectories2}. First define segment $\gamma_{II}$, which has the trajectory
\begin{align}
    \gamma_{II} = (t_{II}(\lambda),x_{II}(\lambda),z_{II}(\lambda))  = (t,t,z_{max}(t_*)) \,\,\,\,\,\,\, -t_* < t < t_*
\end{align}
where $t_*$ is determined by solving $t = |x|_{max}(t)$. By condition 1), this equation will have a solution at finite $t$. By placing this curve at $z(t_*)_{max}$, we ensure that this segment never intersects the brane. 

Next, we construct segments $\gamma_I$ and $\gamma_{III}$. These have the trajectories,
\begin{align}
    \gamma_{I} &= (-t_*+\lambda,-t_*+(1-\epsilon/2)\lambda,z_{max}(t^*)-\sqrt{\epsilon}\lambda) \,\,\,\,\,\,\,\, z_{max}(t^*)/\sqrt{\epsilon} < \lambda < 0 \nonumber \\
    \gamma_{III} &= (t_*+\lambda,t_*+(1-\epsilon/2)\lambda,z_{max}(t^*)-\sqrt{\epsilon}\lambda) \,\,\,\,\,\,\,\,\,\,\,\,\,\,\,\,\,\, 0 < \lambda < z_{max}(t^*)/\sqrt{\epsilon}
\end{align}
It is straightforward to check that these are null, attach to $\gamma_{II}$ when $\lambda=0$, and reach the boundary at finite time. Additionally, by condition 1. they have $x_I(t)<|x|_{max}(t)$ and $x_{III}>|x|_{max}(t)$, so that $\gamma_I$ and $\gamma_{III}$ do not intersect the brane. Thus, the causal curve formed by $\gamma_I\cup\gamma_{II}\cup \gamma_{III}$ traverses from the left to the right CFTs, and shows the bulk geometry is pleasantly traversable. In the next section we establish the two conditions hold for solutions given by perturbing interface to an ellipse elongated in the Euclidean time direction.

\subsubsection*{Shape perturbations of interface brane solutions}

We will now determine explicitly the perturbations of the interface solutions that arise by deforming the CFT interface as shown in figure \ref{fig:perturbedPIc}.

We have seen that the unperturbed interface may be described using Poincar\'e coordinates for the two regions as\footnote{We emphasize that while we use the same letters to describe the coordinates in the two regions, the coordinates generally do not agree at the interface.} 
\be
(z - \gamma_i)^2 + x^2 + \tau^2 = \rho_i^2 \qquad \rho_i \equiv \sqrt{1 + \gamma_i^2} \;
\ee
where $\gamma_i$ ($i=1,2$) is defined in (\ref{defGamma}). We note that
\be
\label{Lrho}
L_1 \rho_1 = L_2 \rho_2 = \frac{1}{T} (\gamma_1 - \gamma_2)
\ee

To consider the perturbations, it will be convenient to make use of shifted spherical coordinates related to the Poincar\'e coordinates as
\be
z - \gamma_i = \rho \cos \theta \qquad \tau = \rho \sin \theta \cos \phi \qquad x = \rho \sin \theta \sin \phi \; .
\ee
Then the metrics become
\be
ds^2 = \frac{L_i^2}{(\rho \cos \theta + \gamma_i)^2} \left[ d \rho^2 + \rho^2 d \theta^2 + \rho^2 \sin^2 \theta d \phi^2 \right]
\ee
and the unperturbed interface is described by $\rho = \rho_i$, with the region $\rho > \rho_1$ in the first spacetime glued to the region $\rho < \rho_2$ in the second spacetime. While the $\phi$ coordinates can be identified in the two patches, the $\theta$ coordinates does not match across the interface. It is convenient to define a coordinate
\be
\tan {\frac{\theta}{2}} = \frac{1}{\rho_i - \gamma_i} \tanh {\frac{g}{2}} .
\ee
With this definition, the induced metrics become
\be
h^{(0)} = L_i^2 \rho_i^2 (dg^2 + \sinh^2 g \,d \phi^2) \; .
\ee
and these agree by the relation (\ref{Lrho}).

The extrinsic curvatures may be calculated as
\be
K_i^{(0)} = {L_i \gamma_i \rho_i \over (\rho \cos \theta + \gamma_i)^2} \left[d \theta^2 + \sin^2 \theta d \phi^2 \right]
\ee
which becomes
\be
K_i^{(0)} = L_i \rho_i \gamma_i  (dg^2 + \sinh^2 g d \phi^2) \; .
\ee
Using the relation (\ref{Lrho}), we can verify the second junction condition
\be
K_1^{(0)} - K_2^{(0)} = T h^{(0)} \; .
\ee

\subsubsection*{First order perturbation}

We now consider a general perturbation
\be
\rho = \rho_1 + \epsilon \delta \rho_1 (\theta, \phi) \qquad \rho = \rho_2 + \epsilon \delta \rho_2 (\theta, \phi)
\ee
of the interface in the two regions. We will solve the junction conditions perturbatively for $\delta \rho$. Other references discussing perturbation theory for the Israel junction conditions include \cite{mukohyama2000perturbation,kodama2000brane}. 

In this case, the induced metrics may be written to first order in $\epsilon$ as
\be
ds^2 = {L_i^2 \rho_i^2 \over (\rho_i \cos \theta + \gamma_i)^2} \left[d \theta^2 + \sin^2 \theta d \phi^2 \right] \left\{ 1+ \epsilon {2 \gamma_i \over \rho_i (\cos \theta \rho_i + \gamma_i)} \delta \rho_i(\theta,\phi)\right\}
\ee
and we find that the extrinsic curvature tensor to first order in $\epsilon$ is $K = K^{(0)} + \epsilon K^{(1)} + \cdots$ where
\be
K^{(0)} = {L_i \gamma_i \rho_i \over (\rho \cos \theta + \gamma_i)^2} \left[d \theta^2 + \sin^2 \theta d \phi^2 \right]
\ee
and
\beas
K^{(1)}_{\theta \theta} &=& {L_i \over (\rho_i \cos \theta + \gamma_i)}\left( - \partial_\theta^2 \delta \rho_i - {\rho_i \sin \theta \over (\rho_i \cos \theta + \gamma_i)} \partial_\theta \delta \rho_i  - \gamma_i { \rho_i \cos \theta - \gamma_i \over  (\rho_i \cos \theta + \gamma_i)^2 } \delta \rho_i \right) \cr
K^{(1)}_{\theta \phi} &=& {L_i \over (\rho_i \cos \theta + \gamma_i)} \left( - \partial_\phi \partial_\theta \delta \rho_i  + \cot \theta \partial_\phi \delta \rho_i  \right) \cr
K^{(1)}_{\phi \phi}&=&{L_i \over (\rho_i \cos \theta + \gamma_i)} \left( - \partial_\phi^2 \delta \rho_i - \sin \theta{\rho_i + \gamma_i\cos \theta \over (\rho_i \cos \theta + \gamma_i)} \partial_\theta \delta \rho_i  - \gamma_i \sin^2 \theta { \rho_i \cos \theta - \gamma_i \over  (\rho_i \cos \theta + \gamma_i)^2 } \delta \rho_i \right)
\eeas
Taking the same transformation to the $g, \phi$ coordinates as above, we find that the induced metrics are
\be
ds^2 = L_i^2 \rho_i^2 \left[d g_i^2 + \sinh^2 g_i d \phi_i^2 \right] \left\{ 1+ \epsilon \gamma_i f_i(g_i,\phi_i) \right\}
\ee
where we define
\be
f_i(g_i,\phi) = {2 \over \rho_i } (\rho_i\cosh g_i - \gamma_i) \delta \rho_i(g_i,\phi_i) \; .
\ee
We have included an index on $g$ and $\phi$ since the coordinates generally do not match between the two surfaces once we perturb them. However, since the coordinates match at leading order, if the induced geometries are the same at first order, we can find a coordinate transformation 
\begin{align}
\label{eq:gphitform}
g_2 &= g_1 - {1 \over 2} \epsilon A(g_1,\phi_1) \cr
\phi_2 &= \phi_1 - {1 \over 2} \epsilon B(g_1, \phi_1)
\end{align}
for which the induced metrics match at first order. Demanding that the first order induced metrics agree gives the equations
\beas
\gamma_2 f_2 - \gamma_1 f_1 &=& \partial_g A \cr
\gamma_2 f_2 - \gamma_1 f_1 &=& \partial_\phi B + \coth (g) A \cr
0 &=& \partial_\phi A + \sinh^2(g)  \partial_g B 
\eeas
Eliminating $f_i$ and $B$, we obtain
\be
\partial_g^2 A - \coth(g) \partial_g A + {\rm csch}^2 (g)  (A + \partial_\phi^2 A) = 0
\ee
Separating variables, we find solutions
\be
A(g,\phi) = e^{i n \phi} \sinh(g) \tanh^{\pm n}\left( {g \over 2} \right) \; .
\ee
For regularity at $g=0$, we should take the positive power of $\tanh$. Restricting also to the real combinations with $\phi \leftrightarrow -\phi$ symmetry, we have solutions
\be
\label{Asol}
A_n(g,\phi) = \cos (n \phi) \sinh(g) \tanh^n \left( {g \over 2} \right)  \; ,
\ee
where the solutions for even $n$ will additionally be symmetric under reversal of Euclidean time. We will mainly be interested in the case $n=2$.

Next, we impose the second junction condition. To first order in $\epsilon$, we have that the extrinsic curvatures for the interface in each region is given in the $g, \phi$ coordinates
\be
K = K_i^{(0)} + \frac{\epsilon}{2} L_i \rho_i K_i^{(1)}
\ee
where each $K_i^{(1)}$ is given in the $(g, \phi)$ coordinates and perturbation $f$ for the corresponding region as 
\begin{align}
K^{(1)} &= \left( (1+2\gamma^2) f-\partial_g^2 f \right) dg^2 + 2\left(\partial_\phi f \coth (g)- \partial_\phi \partial_g f\right)d\phi dg \nonumber \\
    & \left(\left(1+2 \gamma ^2\right) f \sinh ^2(g)- \partial_g f \sinh (g)\cosh (g)-\partial_\phi^2 f\right) d\phi^2\nonumber
\end{align}
The coordinate relations (\ref{eq:gphitform}) give additional terms at first order in $\epsilon$ in $K_2$ that come from expressions $K_2^{(0)}$ in the $(g_1, \phi_1)$ coordinates. Taking these into account, we find that the order $\epsilon$ terms in the second junction condition $K_1 - K_2 = T h$ may be expressed completely in terms of $\delta f = f_2-f_1$ as
\begin{align}
    (\delta f)-\partial_g^2 (\delta f) &= 0 \nonumber \\
    \partial_\phi (\delta f) \coth (g)- \partial_\phi\partial_g (\delta f) &= 0 \nonumber \\
    (\delta f)- \partial_g(\delta f) \text{coth} (g)-\text{csch}^2(g) \partial_\phi^2(\delta f) &= 0
\end{align}
where we have used the $0$th order equation to simplify. The general solution is\footnote{We can check that for the cases where the induced metrics are unperturbed from the original solution, these solutions correspond acting with an infinitesimal scaling operation or translations on the original solution.}
\begin{align}\label{eq:deltafsol}
    \delta f \equiv f_2 - f_1 = \alpha \cosh(g) + (a\cos\phi +b \sin \phi)\sinh(g) 
\end{align}
We note that only modes with $n=0,1$ appear in the general solution for $\delta f$. Higher modes can appear in $f_1,f_2$, but these need to match up trivially across the interface.

Combining (\ref{eq:deltafsol}) with $\gamma_2 f_2 - \gamma_1 f_1 = \partial_g A$ and (\ref{Asol}), we find that the most general solution preserving the reflection symmetries in the $\tau$ and $x$ directions is\footnote{Note that for $\gamma_1=\gamma_2=0$, the interior and exterior metrics agree and our calculation reduces to finding minimal area surfaces. Consequently our solutions reproduce the perturbative solutions studied in \cite{allais2015some} when considering shape deformations of Ryu-Takayanagi surfaces. See in particular equations 41-43 of \cite{allais2015some}, whose first order terms agree with ours.}
\beas
f_1 &=& A_1 \cosh(g) + \sum_{k > 0} C_{2k} \cos (2k \phi) (\cosh(g) + 2k) \tanh^{2k} \left( {g \over 2} \right) \cr
f_2 &=& A_2 \cosh(g) + \sum_{k > 0} C_{2k} \cos (2k \phi) (\cosh(g) + 2k) \tanh^{2k} \left( {g \over 2} \right)
\eeas
We are interested in a perturbation that takes the boundary of the interface to an ellipse. In this case, we have 
\be\label{eq:f1solution}
f_1 =  A\cosh(g) + \cos (2 \phi) (\cosh(g) + 2) \tanh^{2} \left( {g \over 2} \right)
\ee
Here, the parameter $A$ will be chosen so that the boundary of the ellipse in Poincar\'e coordinates will have the same width as the original circle. This requires $f_1=0$ at $\phi=\pi/2$ and $g\rightarrow \infty$, so that $A=1$. 

Using the solution (\ref{eq:f1solution}), we can explicitly verify the two conditions of the last section, establishing that these geometries are smoothly traversable. Before doing so, recall that the function $f_1$ is related to the radial perturbation $\delta \rho$ according to
\begin{align}
    \delta \rho_1 = \frac{\rho_1}{2}\frac{f_1}{\rho_1\cosh g-\gamma_1} = a(g) + b(g) \cos(2\phi)
\end{align}
where 
\begin{align}
    a(g) &= \frac{\rho_1}{2}\frac{\cosh(g)}{\rho_1\cosh(g)-\gamma_1} \nonumber \\
    b(g) & = \frac{\rho_1}{2}\frac{\cosh g +2}{\rho_1\cosh g-\gamma_1} \tanh^2\left( \frac{g}{2}\right)
\end{align}
With this perturbation, the new embedding equation for the brane in Poincar\'e coordinates is
\begin{align}
    x^2 + \tau^2 + (z+\gamma)^2 = (\rho_0+\epsilon \delta \rho)^2 
\end{align}
Using $\delta \rho = a+b\cos 2\phi$ and using some trigonometric identities,
\begin{align}\label{eq:pertrubedbraneembedding}
    x^2 + \tau^2 + (z+\gamma)^2 = (\rho_0+\epsilon(a-b))^2 + 4\epsilon b\cos^2\phi
\end{align}
A parameterization of $x,\tau$ and $z$ which solves this is
\begin{align}
    x &= \tilde{\rho}_1\sin\theta \sin\phi  \label{eq:parametricbraneX} \\
    \tau &= (\tilde{\rho}_1+\epsilon c)\sin\theta\cos\phi  \label{eq:parametricbraneT} \\
    z &= \tilde{\rho}_1\cos\theta - \gamma \label{eq:parametricbraneZ}
\end{align}
where
\begin{align}
    c = \frac{2b}{\rho_1+\epsilon(a-b)}\frac{1}{\sin^2\theta}\,\,\,\,\,\,\,\,\,\,\,\,\,\,\,\,\,\tilde{\rho}_1 = \rho_1 + \epsilon(a-b)
\end{align}
As can be verified explicitly, $c(g)> 0$. This ensures that the intersection of the brane with a surface of constant $g$ forms an ellipse which is elongated in the $\tau$ direction. In particular the profiles are
\begin{align}
    x^2+\tau^2/B^2 = \tilde{\rho}_1 \sin^2\theta \,\,\,\,\,\,\,\,\,\,\,B=1+\epsilon \frac{c}{\tilde{\rho}_1}
\end{align}
In the Lorentzian picture this ensures the fixed $g$ profiles are hyperbolas which asymptote to a straight line with slope $1/B<1$. This gives condition 1, that $d|x|_{max}/dt\leq 1-\epsilon$ for $\epsilon>0$. 

Considering condition 2., we need to establish that $z_{max}(t)$ is finite for finite $t$. From (\ref{eq:pertrubedbraneembedding}) we have the bound
\begin{align}
    z \leq \sqrt{\tilde{\rho}_1^2 +4\epsilon b \cos^2\phi + t^2} - \gamma \leq \sqrt{\tilde{\rho}_1^2 +4\epsilon b +t^2}
\end{align}
where we've also gone to Lorentzian time. We may explicitly check that this is bounded (as a function of $g$) for bounded $t$, which establishes condition $2.$

\section{Conclusion}

In this paper, following \cite{VanRaamsdonk:2018zws}, we have investigated in more detail how an alternative set of degrees of freedom --- a collection of BCFTs with or without interactions --- can be used to approximate an AdS spacetime dual to the vacuum state of a single CFT. We have emphasized that without interactions, the causal structure is that of a non-traversable multi-boundary wormhole. The non-traversability is enforced by the dynamics of ETW branes. By adding back interactions in various ways, the ETW-brane trajectories are modified and in some cases, the result is a traversable geometry. 

We emphasize that in all the cases we consider, the interior geometry always includes an arbitrarily large causal patch of pure AdS; the differences in the ETW brane trajectories only affect what happens outside this region.\footnote{In more detailed microscopic examples, the interior geometry may not be precisely the same in the approximated geometries, but we expect that it should approach the original geometry in the limit where the modifications to the Euclidean path integral are small.} So we have many different descriptions of the same interior geometry.

From one perspective, we should not be surprised that an alternative set of degrees of freedom can be used to describe AdS physics. We can, at least plausibly, simulate a holographic CFT on a (quantum or classical) computer. The degrees of freedom that compose the computer are well understood to be unimportant --- a computer may consist of traditional silicon hardware, superconducting qubits, or (more esoterically) colliding billiard balls, etc. Any of these hardwares can be used to build a universal computer, which can then run the simulation of the CFT. What is perhaps more surprising is that the map from the holographic CFT to a set of discrete degrees of freedom can be as simple as the discretization operator given in equation (\ref{eq:discretization}). As well, a traditional simulation does not preserve the local structure of the CFT, in that there is no sense in which a subregion of the CFT is simulated by a subregion of the computer, as occurs here. 

As we mentioned in the introduction, the many possible ``architectures'' for describing the physics of a finite causal patch in quantum gravity suggest that the local gravitational physics is more related to fundamental quantum  information  theoretic  properties of the state rather than  details  about  the  underlying  degrees of  freedom. It will be interesting to understand in more detail which information-theoretic features are essential for this emergence of gravitational physics.\footnote{For a recent discussion, see \cite{may2020holographic}, who argue that $O(1/G_N)$ entanglement between boundary subregions is essential without making reference to the specific boundary theory.} 

\vspace{0.25cm}
\noindent \textbf{Acknowledgements}
\vspace{0.25cm}

We thank Brian Swingle for questions and conversations that led to this project. We also thank Christopher Waddell, David Wakeham, and Petar Simidzija for helpful conversations. AM is supported by a C-GSM award given by the National Science and Engineering Research Council. This research is supported in part by the Natural Sciences and Engineering Research Council of Canada and by the Simons Foundation though a Simons Investigator award and the ``It From Qubit'' Collaboration grant.

\appendix 

\section{Propagators in AdS/BCFT}

\subsection{Bulk-to-bulk and bulk-to-boundary propagator}\label{sec:bulkboundarypropagator}

In this appendix we calculate the two-point function in the presence of a zero tension brane. 

In Euclidean Poincar\'e-AdS space (with no brane present), we have the two-point function \cite{Erdmenger2015}
\begin{align}
    G_\Delta(X,X')=\frac{C_\Delta}{2^\Delta (2\Delta - d)}\eta^\Delta {}_2F_1\left(\frac{\Delta}{2},\frac{\Delta+1}{2};\Delta-\frac{d}{2}+1;\eta^2 \right)
\end{align}
with
\begin{align}
    \eta = \frac{2zz'}{z^2+(z')^2+(x-x')^2+(\tau-\tau')^2}
\end{align}
and
\begin{align}
    C_\Delta = \frac{\Gamma(\Delta)}{\pi^{d/2}\Gamma(\Delta-d/2)}.
\end{align}
To get the Lorentzian two-point function, we should Wick rotate by setting $it=\tau$. We should be careful though not to cross poles or branch cuts in doing so. One may check that the branch points of the propagator occur on the $\tau-\tau'=T$ imaginary axis, so we can rotate by $e^{\pm i(\pi-\epsilon)}$. 

At zero tension in Euclidean space, the brane is just a half sphere in Poincar\'e coordinates
\begin{align}
    x^2+\tau^2 + z^2=1.
\end{align}
We can define spherical coordinates,
\begin{align}
    \tau &= r \sin \theta \sin \phi\\
    x &= r \sin \theta \cos \phi \\
    z &= r\cos \theta
\end{align}
so that the brane is at $r=1$. We may express the chordal distance $\eta$ in terms of spherical coordinates,
\begin{align}
    \eta = \frac{2rr'\cos\theta \cos \theta'}{r^2 + (r')^2- 2rr'\sin\theta \sin\theta' [\cos\phi \cos\phi' +\sin\phi\sin\phi']}.
\end{align}
We consider the two-point function as a function of these coordinates,
\begin{align}
    G_\Delta(r,\theta,\phi;r',\theta',\phi')
\end{align}
The two-point function with Neumann boundary conditions is then easy to guess,
\begin{align}
    G_{\Delta,N}(r,\theta,\phi;r',\theta',\phi') = G_\Delta(r,\theta,\phi;r',\theta',\phi') + G_\Delta(r'r,\theta,\phi;1,\theta',\phi')
\end{align}
One can check explicitly that this satisfies the boundary condition and returns a delta function when acted on with the Klein-Gordon operator. 

Next we calculate the bulk-boundary propagator $\langle \varphi \mathcal{O}\rangle$. We can calculate this using
\begin{align}
    \langle \varphi(X)\mathcal{O}(x')\rangle = (2\Delta -d) \lim_{z'\rightarrow 0} (z')^{-\Delta} G_{\Delta,N}(r,\theta,\phi;r',\theta',\phi') \nonumber 
\end{align}
The resulting bulk to boundary correlator is
\begin{align}\label{eq:bulkboundary}
    \langle \varphi(x,\tau,z)\mathcal{O}(x',\tau')\rangle=\frac{C_\Delta z^\Delta}{[z^2+(x-x')^2+(\tau-\tau')^2]^\Delta} + \frac{C_\Delta z^\Delta}{[1-2(xx'+\tau\tau')+(x^2+\tau^2+z^2)((x')^2+(\tau')^2)]^\Delta}. \nonumber 
\end{align}
The first term is the usual bulk-to-boundary propagator without a brane. 

\subsection{Bulk-to-edge propagator}\label{sec:edgepropagator}

In the main article we need the bulk-edge correlator,
\begin{align}
    \langle \varphi(X)\psi(s) \rangle
\end{align}
where $\psi(s)$ is an edge operator. To get this we can begin with the bulk-boundary correlator, 
\begin{align}
    \langle \varphi(X)\mathcal{O}(x)\rangle
\end{align}
And apply the \emph{boundary operator expansion} (BOE) of $\mathcal{O}$. The BOE expresses a CFT bulk operator in terms of a sum over CFT boundary operators.

For a BCFT on the half plane, the BOE takes the general form,
\begin{align}
    \mathcal{O}_\Delta(x,y) = \sum_k P_{\Delta}^k |y|^{h_k-\Delta} \psi_k(x),
\end{align}
where $y$ is the coordinate transverse to the boundary. In our setting however we need the BOE when the boundary is the unit disk, with the CFT defined on the disk exterior. The expansion should now be taken in terms of the radial coordinate $r$,
\begin{align}\label{eq:radialBOE}
    \mathcal{O}_\Delta(r,\phi) = \sum_k R_{\Delta}^k (r-1)^{h_k-\Delta} \psi_k(\phi).
\end{align}
Using a conformal transformation from the half plane to the disk exterior, it's possible to derive the form of this expansion and relate the coefficients appearing here to those in the half plane expansion. 

We put this expansion into the two point function
\begin{align}\label{eq:bulktoedge}
    \langle \varphi(X)\mathcal{O}(r,\phi)\rangle = \sum_k R_{\Delta}^k (r-1)^{h_k-\Delta} \langle \varphi(X) \psi_k(\phi) \rangle
\end{align}
We will choose to couple the left and right BCFTs using the first operator $\psi_k$ in this expansion. Writing the bulk-to-boundary propagator in a series expansion around $r=1$, we find that the leading term has $h_1=\Delta$. The bulk-edge propagator then is
\begin{align}
    \langle \varphi(X) \psi_k(\phi) \rangle = \frac{1}{R_\Delta^1} \lim_{r\rightarrow 1} \langle \varphi(X)\mathcal{O}(r,\phi)\rangle.
\end{align}
Using expression (\ref{eq:bulkboundary}), we find
\begin{align}
    \langle \varphi(r,\theta,\phi) \psi_1(\phi') \rangle = \frac{2\tilde{C}_\Delta z^{\Delta}}{[1+r^2-2r\cos(\phi-\phi')\sin\theta]^{\Delta}}\,, 
\end{align}
where we defined a new constant $\bar{C}_{\Delta}\equiv C_\Delta / R_{\Delta}^k$. 

We will fix the normalization of the bulk-edge propagator by requiring the edge-edge correlation function be unit normalized. In particular, taking the limit of the bulk operator to the edge yields the edge-edge correlation function
\begin{align}
    \langle\psi_1(\phi)\psi_1(\phi') \rangle &= \lim_{r\rightarrow 1} \lim_{z\rightarrow 0} z^{-\Delta} \langle \varphi(r,\theta,\phi) \psi_1(\phi') \rangle \\
    &= \frac{1}{[1-\cos(\phi-\phi')]^\Delta}
\end{align}
which requires we fix $2\tilde{C}_{\Delta}=2^\Delta$.

In Poincar\'e coordinates, and fixing the normalization, the bulk-edge correlation function is
\begin{align}
    \langle \varphi(z,x,\tau) \psi_1(\phi) \rangle = \frac{2^\Delta z^\Delta}{[1+\tilde{r}^2-2x\cos (\phi) - 2\tau \sin(\phi)]^{\Delta}} 
\end{align}
where $r^2 = x^2+z^2 + t^2$. Finally we can Wick rotate $\phi\rightarrow is$ to obtain the Lorentzian bulk-to-edge propagator,
\begin{align}
    \langle \varphi(z,x,t) \psi(s) \rangle = \frac{2^\Delta z^\Delta}{[1+\tilde{r}^2-2x\cosh(s) + 2t\sinh(s)-i\epsilon]^{\Delta}}.
\end{align}

\section{Stress tensor calculation}\label{sec:stresstensorcalc}

We can use the functional forms for the bulk-to-edge propagator to write out the leading correction term of the two-point function explicitly, 
\begin{align}
    G_h &= h 2^{2\Delta+1} \sin(\pi \Delta) z^{2\Delta}\times \nonumber \\
    &\,\,\,\,\,\,\int_{-\infty}^{+\infty} ds \left[ \theta(s)\theta(ue^{-s}-1-z^2)  \frac{1}{(1+z^2+u'e^s)^{\Delta}} \frac{1}{|1+z^2-ue^{-s}|^{\Delta}} + (u\leftrightarrow u')\right]. \nonumber
\end{align}
In the second line we've set $x=t$. Because equation (\ref{eq:tuulimit}) is symmetric in $u\leftrightarrow u'$, the second term in the above integral gives the same contribution to $T_{uu}$ as the first, and we can include it by adding a factor of $2$. We use the theta functions to restrict the range of integration,
\begin{align}
    G_h= h 2^{2\Delta+2} \sin(\pi \Delta) z^{2\Delta} \theta(u-1-z^2) \int_{0}^{\ln \left(\frac{u}{1+z^2}\right)} ds   \frac{1}{(1+z^2+u'e^s)^{\Delta}} \frac{1}{(ue^{-s}-1-z^2)^{\Delta}}. \nonumber
\end{align}
The $\theta(u-1-z^2)$ appears because $\theta(s)\theta(1+z^2-ue^{-s})$ is zero for all $s$ unless $u>1+z^2$. 

Now we take the $\partial_{u'}$ derivative. 
\begin{align}
    \partial_{u'}G_h = -\Delta h 2^{2\Delta+2} \sin(\pi \Delta) z^{2\Delta} \theta(u-1-z^2) \int_{0}^{\ln \left(\frac{u}{1+z^2}\right)} ds\,e^{s}  \frac{1}{(1+z^2+u'e^s)^{\Delta+1}} \frac{1}{(ue^{-s}-1-z^2)^{\Delta}}, \nonumber
\end{align}
then substitute $x=e^s$,
\begin{align}
    \partial_{u'}G_h = - \Delta h 2^{2\Delta+2} \sin(\pi \Delta) z^{2\Delta} \theta(u-1-z^2) \int_{1}^{ \left(\frac{u}{1+z^2}\right)} dx  \frac{1}{(1+z^2+u'x)^{\Delta+1}} \frac{1}{(u/x-1-z^2)^{\Delta}} . \nonumber
\end{align}
Now we re-scale the $u$ coordinate, $u=\tilde{u}(1+z^2)/2$, and the $u'$ coordinate in the same way,
\begin{align}
    \partial_{u'}G_h = -\Delta h  \sin(\pi \Delta) \frac{2^{4\Delta +3}z^{2\Delta}}{(1+z^2)^{2\Delta+1}}\theta(\tilde{u}-2) \int_{1}^{ \left(\frac{\tilde{u}}{2}\right)} dx \frac{1}{(2+\tilde{u}'x)^{\Delta+1}} \frac{1}{(\tilde{u}/x-2)^{\Delta}} . \nonumber
\end{align}
The integral now has its $z$ dependence removed, and in fact corresponds to setting $z=1$ in the original integral. We will collect the $z$ dependence of the pre-factor into a function $f(z,\Delta)$, so that
\begin{align}
    \partial_{u'}G_h = -h f(z,\Delta) \theta(\tilde{u}-2) \int_{1}^{ \left(\frac{\tilde{u}}{2}\right)} dx  \frac{1}{(2+\tilde{u}'x)^{\Delta+1}} \frac{1}{(\tilde{u}/x-2)^{\Delta}} \equiv \kappa(u,u',z), \nonumber
\end{align}
where we defined $\kappa$ in the last line, and
\begin{align}
    f(z,\Delta)=\Delta \sin(\pi \Delta) \frac{2^{4\Delta +3}z^{2\Delta}}{(1+z^2)^{2\Delta+1}}.
\end{align}
Recall that our eventual goal is to calculate
\begin{align}
    \int_{-\infty}^{+\infty}du \,T_{uu}(u,z) = \int_{-\infty}^{+\infty}du \,\lim_{u\rightarrow u'}\partial_u\partial_u' G_h(u,u',z)=\int_{-\infty}^{+\infty}du \,\lim_{u\rightarrow u'}\partial_u\kappa(u,u',z).
\end{align}
We'll note that
\begin{align}\label{eq:kappatricks}
    \lim_{u\rightarrow {u'}}\partial_u\kappa(u,u',z) = \partial_u [\kappa(u,u,z)] - \lim_{u'\rightarrow u}\partial_{u'}[\kappa(u,u',z)].
\end{align}
Then, we can show that the du integral over the first term vanishes. To see this, we use the change of variables
\begin{align}\label{xtoycoordchange}
    y = \frac{x-1}{\tilde{u}/2-1}
\end{align}
to do the integral appearing in $\kappa$ explicitly. We find
\begin{align}
    \kappa(\tilde{u},\tilde{u},z) = 4h\frac{f(z,\Delta)}{1-\Delta}\theta(\tilde{u}-2)\frac{ \tilde{u}^\Delta(\tilde{u}-2)^{1-\Delta}}{(4+\tilde{u}^2)^{\Delta+1}}F_1[1-\Delta,-\Delta,1+\Delta,2-\Delta,\frac{\tilde{u}-2}{\tilde{u}},\frac{\tilde{u}(\tilde{u}-2)}{4+\tilde{u}^2}], \nonumber 
\end{align}
where $F_1$ is the Appell hypergeometric function. The $\theta$ function reduces the integral over $du$ to the interval $\tilde{u}\in[2,\infty)$. We can then see that $\kappa(u,u,z)$ vanishes for $u\in\{2,\infty\}$, so that indeed the first term in (\ref{eq:kappatricks}) doesn't contribute to $\int du T_{uu}$. 

Keeping only the second term then, we have
\begin{align}
    \int_{-\infty}^{+\infty}du \,T_{uu}(u,z) &=-\int_{-\infty}^{+\infty}du \,\lim_{u\rightarrow u'}\partial_{u'}\kappa(u,u',z) \nonumber \\
    &=  -2h f(z,\Delta)(\Delta+1) \int_{2}^{+\infty} d\tilde{u}\int_{1}^{ \left(\frac{\tilde{u}}{2}\right)} dx  \frac{x}{(2+\tilde{u}x)^{\Delta+2}} \frac{1}{(\tilde{u}/x-2)^{\Delta}}
\end{align}
again using the coordinate transformation \ref{xtoycoordchange}, this is
\begin{align}
    \int_{-\infty}^{+\infty}du \,T_{uu}(u,z) = -2h \frac{f(z,\Delta)(1+\Delta)}{(1-\Delta)} &\int_{2}^{\infty}d\tilde{u}\frac{\tilde{u}^{1+\Delta}(\tilde{u}-2)^{1-\Delta}}{(4+\tilde{u}^2)^{\Delta+2}} \nonumber \\
    & \times F_{1}[1-\Delta,-1-\Delta,2+\Delta,2-\Delta,\frac{\tilde{u}-2}{\tilde{u}},\frac{\tilde{u}(\tilde{u}-2)}{4+\tilde{u}^2}].\nonumber 
\end{align}
Finally, we insert this into expression (\ref{eq:traversabilitycondition}) for $v(\infty)$, and find
\begin{align}\label{eq:explicitvinfty}
    v(\infty)=-16\pi G_N h\left(\int_{+\infty}^{z} dz' z' \int_{+\infty}^{z'} \frac{dz''}{z''} F[z'',\Delta]\right)\left(\int_{2}^{+\infty}d\tilde{u} \, H(\tilde{u},\Delta)\right),
\end{align}
where we defined the functions
\begin{align}
    F[z,\Delta]&=2 f(z,\Delta)\frac{(1+\Delta)}{(1-\Delta)}, \nonumber \\
    H[\tilde{u},\Delta]&=\frac{\tilde{u}^{1+\Delta}(\tilde{u}-2)^{1-\Delta}}{(4+\tilde{u}^2)^{\Delta+2}} F_{1}[1-\Delta,-1-\Delta,2+\Delta,2-\Delta,\frac{\tilde{u}-2}{\tilde{u}},\frac{\tilde{u}(\tilde{u}-2)}{4+\tilde{u}^2}].
\end{align}
The integration over $F[z,\Delta]$ can be done explicitly, 
\begin{align}\label{eq:Fz}
    \left(\int_{+\infty}^{z} dz' z' \int_{+\infty}^{z'} \frac{dz''}{z''} F[z'',\Delta]\right) = \frac{4^{2 \Delta +1} \sin (\pi  \Delta ) \, _2F_1\left(\Delta ,2 \Delta +1;\Delta +2;-\frac{1}{z^2}\right)}{z^{2\Delta}(1-\Delta)}
\end{align}
The integral over $H[\tilde{u},\Delta]$ we do numerically. The final result is a plot of $v(\infty)$, which we can explore as a function of $\Delta$ or of $z$. The $z$ dependence is controlled by (\ref{eq:Fz}), while we show $v(\infty)$ as a function of the operator weight $\Delta$ with $z=1$ in figure \ref{fig:deltavplot}. Note that $v<0$ for all values of $\Delta$ such that $0<\Delta\leq 1$. As well, $v(\infty)=0$ when $\Delta=0$. This is expected, as $\Delta=0$ corresponds to inserting the identity operator.

\section{Backreaction on the brane}
\label{sec:backreaction}

In section \ref{sec:stresstensorcorrection}, we focused on finding the trajectory of a null ray inside the brane. We found that with the coupling term turned on, one can find a null ray which travels from one CFT boundary to the other, rendering the geometry traversable. Conveniently, we found that even first order effects on the brane trajectory could be ignored in this calculation.

In this section, we would like to understand the effects of our perturbation on the ETW brane geometry itself. We investigate the perturbation of the brane trajectory in Fefferman-Graham coordinates, focusing on the light ray $v=0$, $z=1$ which lies inside the ETW brane in the unperturbed geometry. We will find that the brane trajectory along $v=0$ moves toward smaller $z$ for large $u$, indicating that the ETW brane expands more slowly into the $z$ direction when the coupling is present. 

Recall that the brane position is fixed by the boundary condition (\ref{eq:ADSBCFTbc}), which we specialize to the $T=0$ case,
\begin{align}\label{eq:bc}
    K_{ab} - K h_{ab}= 0.
\end{align}
We'd like to consider this boundary condition perturbatively. In particular we consider a metric perturbation,
\begin{align}\label{eq:metric}
    ds^2 = \frac{L^2}{z^2}(dudv + dz^2 + \epsilon \delta g_{\mu\nu}dx^\mu dx^\nu),
\end{align}
which results in some surface deformation $f(u,v)$
\begin{align}\label{eq:embed}
    0 = uv + z^2 - 1 + \epsilon f(u,v) ,
\end{align}
where we've used light cone coordinates $u=x+t$, $v=x-t$. We consider $u,v$ as the intrinsic coordinates on the brane, and consider $z$ to be a function of $u,v$. It is then routine to insert this metric and surface and calculate the induced stress tensor according to (\ref{eq:bc}). 

Following the traversability calculation in section \ref{sec:doubletrace}, we go to the $v=0$ ray along the brane, where also $z=1+O(\epsilon)$. We get the three components of the boundary condition, the $uu$ component of which is simple 
\begin{align}\label{eq:uu}
    0 &= \delta g_{uu}+ \partial_u(u\delta g_{uu}) - z \partial_z \delta g_{uu}  - \partial_u^2f\,.
\end{align}
Conveniently, this involves only the function $f$ we are interested in along with the $\delta g_{uu}$ term of the metric perturbation.

Next, integrate over the $v=0$ ray, yielding
\begin{align}
     \left[(1-\partial_z)\int_{-\infty}^{u_0}du \,\delta g_{uu}(z) \right]_{z=1} + u_0 \delta g_{uu}(u_0) = \partial_u f(u)|_{u=u_0}.
\end{align}
Note that we have used $f=0$ at early times, before the perturbation is turned on, and used that $u\delta g_{uu}(u)\rightarrow 0$ when $u\rightarrow-\infty$. Next, note that for large enough $u_0$ the integrated term above dominates over the $u_0 \delta g_{uu}(u_0)$ term. This is because $u\delta g_{uu}(u)\rightarrow 0$ when $u\rightarrow \infty$, while the first term remains finite. Given this, we'll drop this second term in what follows.  

From the embedding equation (\ref{eq:embed}), we can see that $f$ controls the $z$ position of the brane, 
\begin{align}
    \epsilon \partial_uf(u,0) = -2\partial_u z(u,0).
\end{align}
Thus we find
\begin{align}\label{eq:partialzandANEC}
    \epsilon \left[(1-\partial_z)\int_{-\infty}^{u_0}du \,\delta g_{uu}(z) \right]_{z=1} = -2\partial_u z(u=u_0)
\end{align}
Recalling (\ref{eq:traversabilitycondition}) and our conclusion that the deformation results in $v(\infty)<0$, we have
\begin{align}
    -v(\infty)=\int_{-\infty}^{+\infty} du\,\delta g_{uu} > 0
\end{align}
Since $\delta g_{uu}\rightarrow 0$ as $u\rightarrow \infty$, this will also be positive for a sufficiently large but finite upper bound on the $du$ integral. Finally, we would like to know the $z$ dependence of this integral. This is controlled by (\ref{eq:explicitvinfty}) and (\ref{eq:Fz}), which give that the $z$ derivative is negative, so that
\begin{align}
    -\partial_z \int_{-\infty}^{u_0}du \,\delta g_{uu}(z) >0.
\end{align}
Using these two inequalities in (\ref{eq:partialzandANEC}), we find that for $u_0$ large $\partial_u f<0$, that is the brane is moving towards the boundary (relative to its initial position) along the $v=0$ light ray. While we were not able to solve for the full perturbed brane position, this result suggests the brane expands more slowly in the radial direction by the presence of the coupling.

\bibliographystyle{JHEP}
\bibliography{biblio}

\end{document}